\documentclass[aps,showpacs,11pt]{revtex4}
\usepackage{dcolumn}
\usepackage{graphicx}
\usepackage{amsmath}
\usepackage{amsfonts}
\usepackage{amssymb}
\usepackage{psfrag}
\usepackage{wrapfig}
\usepackage{subfigure}
\usepackage{makeidx}
\usepackage{bm}
\usepackage{epsf}
\usepackage{hyperref}
\usepackage{color}
\usepackage{multirow}

\begin{document}
	
	\title{Object picture of scalar field perturbation on Kerr black hole in scalar-Einstein-Gauss-Bonnet theory}
	
	\author{Shao-Jun Zhang$^{1,2}$}
	\email{sjzhang84@hotmail.com}
	\author{Bin Wang$^{3,4}$}
	\email{wang$_$b@sjtu.edu.cn}
	\author{Anzhong Wang$^{5}$}
	\email{Anzhong$_$Wang@baylor.edu}
	\author{Joel F. Saavedra$^{6}$}
	\email{joel.saavedra@pucv.cl}
	\affiliation{$^1$Institute for Theoretical Physics $\&$ Cosmology, Zhejiang University of Technology, Hangzhou 310032, China\\
		$^2$United Center for Gravitational Wave Physics, Zhejiang University of Technology, Hangzhou 310032, China\\
		$^3$Center for Gravitation and Cosmology, Yangzhou University, Yangzhou 225009, China
		\\ $^4$School of Aeronautics and Astronautics, Shanghai Jiao Tong University, Shanghai 200240, China\\
		$^5$GCAP-CASPER, Physics Department, Baylor University, Waco, TX 76798-7316, USA\\
		$^6$Instituto de F\'isica, Pontificia Universidad Cat\'olica de Valpara\'iso, Casilla 4950, Valpara\'iso, Chile}
	\date{\today}

	\begin{abstract}
		\indent Scalar perturbations around the Kerr black hole in scalar-Einstein-Gauss-Bonnet (sEGB) theory are studied in the time domain. To overcome the ``outer boundary problem" that usually encountered in traditional numerical calculations, we apply the hyperboloidal compactification technique to perform a $(2+1)$-dimensional simulation aiming to obtain a precise object picture of the wave propagation under the scalar field perturbation. We find that the big enough coupling constant between the scalar field and the Gauss-Bonnet curvature is responsible to destroy the original Kerr black hole. The breakdown of the Kerr spacetime happens earlier and the instability becomes more violent when the coupling becomes stronger.  We further present object confirmations on the special case for the negative coupling where there exists a minimum rotation and below which the instability can never happen no matter how strong the coupling is. We also illustrate the fine structure property in the quasinormal ringing frequency once there is the coupling, and present the characteristic imprint of the sEGB theory. We expect that such a fine structure  can be detected in the future gravitational wave observation to test the sEGB theory.
		
	\end{abstract}
	

	\maketitle
	
	\section{Introduction}
	
	In the past few years, because of the great achievements in observations, especially the first detection of the gravitational wave(GW) \cite{Abbott:2016blz,Abbott:2016nmj,Abbott:2017gyy} and  the first-ever picture captured for the black hole  \cite{Akiyama:2019cqa,Akiyama:2019eap}, the study of black hole physics has entered the golden era. These observations enable us to test general relativity (GR) in  the strong gravity regime \cite{Carter:1971zc,Robinson:1975bv,Chrusciel:2012jk}. In GR,  the Kerr metric describes the most general black hole spacetime (without charge). Besides mass and rotation, the Kerr black hole does not have additional hairs, which is the famous no-hair theorem. However this no-hair theorem is challenged in the modified gravity theories (MOGs), where the general hairy rotating black hole solutions were found \cite{Sotiriou:2015pka,Herdeiro:2015waa}. More interestingly it was found that the  Kerr  solution can also be allowed in MOGs, although in MOGs the perturbation on the same Kerr 
	background is in general different from that in GR. This actually gives a  possible way to distinguish MOGs from GR though the study of perturbation dynamics.  Employing the current  available observational constraints, some MOGs, such as scalar-tensor theory, scalar-Einstein-Gauss-Bonnet theory (sEGB), dynamical Chern-Simon gravity (dCSG) and Lorentz-violating gravity (see, e.g., \cite{Berti:2018cxi} for a review) were found viable. It is intriguing to further disclose the signature of MOGs by uncovering the objective picture of wave dynamics. This in turn can check the no-hair theorem and unveil the hairy black holes imprints. Precise pictures in wave dynamics around hairy black holes can help understand better physics in MOGs through future GW detections.

	Among MOGs, the sEGB theory has recently attracted a lot of attentions. In this theory, an additional scalar field coupling to the Gauss-Bonnet (GB) term is added to the usual Einstein-Hilbert action. Spherical and axial symmetric black hole solutions in GR were found in the sEGB theory \cite{Antoniou:2017acq,Doneva:2017bvd,Silva:2017uqg,Cunha:2019dwb,Herdeiro:2020wei,Berti:2020kgk}. However, when the coupling between the scalar field and GB term becomes strong enough, the effective mass square can become negative in the vicinity of the black hole horizon, resulting in the tachyonic instability which leads to the so-called spontaneous scalarization and the formation of the hairy black holes.  The phenomenon of spontaneous scalarization was observed a long time ago in neutron stars, but there the instability is induced by the surrounding matter instead of the curvature \cite{Damour:1993hw}. Hairy black holes show the signature of MOGs. A lot of efforts have been devoted to study the physical implications of these hairy black holes, including black hole shadows \cite{Cunha:2019dwb}, the classical stability \cite{Blazquez-Salcedo:2018jnn,Silva:2018qhn},  etc. Results that show deviations from that of the Kerr solution indicate possible observational signatures to test sEGB theory in the future.

	Perturbations around black holes can reflect characteristic sounds of black holes which can be recorded by GW detectors. Studying matter perturbations such as the scalar field perturbations in the black hole background has both theoretical and observational interests. Theoretically, it has been known for a long time that the time evolution of the perturbations in the black hole background will generally experience three stages \cite{Vishveshwara:1970zz}. After the initial pulse, the perturbation field undergoes the damped oscillations,  called  the quasinormal ringing with frequencies and damping times  entirely fixed by the black hole parameters.  At late times, quasinormal oscillations are swamped by the relaxation process, which is the requirement of the black hole no-hair theorem. For more details, please refer to the reviews \cite{Kokkotas:1999bd,Nollert:1999ji,Berti:2009kk,Konoplya:2011qq}. The coupling between  the scalar field and GB term must influence the quasinormal modes (QNMs) and  their late time wave dynamics.  When the coupling is strong enough, the QNMs can show its effect and the late time wave behavior can reflect the destroy of the GR black hole,   because of the tachyonic instability.  The blow up of the perturbation tail illustrates the breakdown of the no-hair theorem and the birth of the hairy black hole. In observations, through GW detections, the MOGs can be confirmed if the signatures of the wave dynamics in sEGB theory are detected. It is expected that current and further GW detectors will detect the signatures of the ringdown and late-time stages \cite{Berti:2018vdi}.    
	
	In GR, even for the most general Kerr background, thanks to the separability of the scalar perturbation equation, the QNMs have been extensively studied either in the frequency domain or time domain, for references on this topic please see \cite{Kokkotas:1999bd,Nollert:1999ji,Berti:2009kk,Konoplya:2011qq}. However, the situation becomes more complicated in sEGB, since the scalar perturbation equation cannot be separated with  usual methods. Because of this reason, in the study of the spontaneously scalarized Kerr black holes, the objective picture of perturbation wave dynamics was not presented explicitly in \cite{Cunha:2019dwb} when the coupling between the scalar field and GB curvature correction term $\alpha>0$. 
	This technical obstacle is common in MOGs. Fortunately, several methods, mostly numerical, have been developed in the past few years to deal with such problems  \cite{Krivan:1996da,PazosAvalos:2004rp,Dolan:2011dx,Thuestad:2017ngu,Dolan:2012yt,Brito:2014nja,Zenginoglu:2007jw,Zenginoglu:2008pw,Zenginoglu:2008uc,Zenginoglu:2008wc,Zenginoglu:2009ey,Zenginoglu:2009hd,Zenginoglu:2010cq,Zenginoglu:2011jz,Racz:2011qu,Zenginoglu:2011zz,Cano:2020cao}. Among them, a general method is to study the full time evolution of the scalar perturbation by a $(1+1)$ or $(2+1)$-dimensional simulation. In  \cite{Dima:2020yac}, by projecting the scalar field onto a basis set of spherical harmonics to separate the angular dependencies, the authors considered the case with $\alpha<0$ and studied the time evolution of the scalar perturbation using  $(1+1)$-dimensional simulations. The results show that instability only occurs at sufficiently high spins $(a/M > \frac{1}{2})$ (see also Ref. \cite{Hod:2020jjy} for an analytical argument). For a  given big enough spin $a$, there exists a critical value $|\alpha_c|$, above which instability occurs. However,  $|\alpha_c|$ depends on the parameter $a$, and in particular
	it decreases as $a$ is increasing.  Further, these results were confirmed by using the $(2+1)$-dimensional simulation with a numerical method introduced in \cite{Doneva:2020nbb}. It was shown that the instability with $\alpha<0$ leads to the dubbed spin-induced scalarization and the final hairy black holes can be constructed \cite{Herdeiro:2020wei,Berti:2020kgk}. See also a related work Ref. \cite{Konoplya:2019hml} concerning the time evolution of a test and minimally coupled scalar field in this theory.
	
	In available numerical treatments, the tortoise coordinate was utilized to map the radial computational domain $(r_+, +\infty)$ to $(-\infty, +\infty)$ (here $r_+$ is the horizon). However, Price {\it et al}. pointed out that traditional numerical methods suffer the so-called ``outer boundary problem" \cite{Thuestad:2017ngu}: namely in practical calculations, one has to truncate the infinite radial computational domain to a finite range and put boundary conditions (ingoing and outgoing waves at the horizon and infinity,
	 respectively) at the outer edges, thus inevitably resulting spurious wave reflections from the edge which can spoil the evolution of the scalar perturbation at the late time. To overcome this problem, a technique called hyperboloidal compactification was developed in \cite{Zenginoglu:2007jw,Zenginoglu:2008pw,Zenginoglu:2008uc,Zenginoglu:2008wc,Zenginoglu:2009ey,Zenginoglu:2009hd,Zenginoglu:2010cq,Zenginoglu:2011jz,Zenginoglu:2011zz}, with which the radial computation domain is mapped to a finite range and the outer boundary conditions are satisfied automatically. Interestingly, armed with this numerical strategy, only until recently we obtain a well understanding on the late-time behaviors of  the scalar field  perturbation even for 
	 the Kerr black hole \cite{Zenginoglu:2012us,Burko:2013bra,Thuestad:2017ngu}. This technique has also been successfully applied to study gravitational waveforms from large- and extreme-mass-ratio inspirals in 
	  the  Kerr black hole extracted at null infinity for the first time \cite{Zenginoglu:2011zz}, and also the full-time evolution of scalar perturbations in dCSG \cite{Gao:2018acg}. 
	
	It is interesting to further apply the hyperboloidal compactification technique to study the scalar perturbation in the sEGB theory when there is coupling between the scalar field and GB curvature correction.  We expect to get a complete and objective picture on the wave dynamics to describe precisely the time evolution of the scalar perturbation in the sEGB theory. Such a picture can not only provide us information about the stability of the black hole, serving as an independent confirmation of the scalarization discussions \cite{Cunha:2019dwb, Dima:2020yac, Doneva:2020nbb,Hod:2020jjy}, but also it can present further imprints on the scalar dynamics, including QNMs and the late-time behavior influenced by the coupling constant between the scalar field and GB term. Precise waveforms can reflect the effect of the coupling $\alpha$ in QNMs and suggest the potential verification of the sEGB theory.  We will examine the objective picture of the scalar evolution with $\alpha>0$ as well as $\alpha<0$ in the time domain,  and obtain a complete understanding on the effects of the coupling constant on the scalar wave dynamics.

	This paper is organized as follows. In Sec. II, we give a brief introduction of the sEGB theory and write out the scalar perturbation equation. In Sec. III, applying the hyperboloidal compactification technique, we cast the scalar perturbation equation into a form suitable for numerical calculations. In Sec. IV, we first check the validity of the hyperboloidal compactification technique for the case $\alpha=0$, and then we report numerical results for $\alpha>0$ and $\alpha<0$, respectively, in sEGB theory. The last section is devoted to summary and discussions.

	\section{Scalar-Einstein-Gauss-Bonnet theory and scalar perturbation equation}
	
	The action of sEGB theory is \cite{Antoniou:2017acq,Doneva:2017bvd,Silva:2017uqg,Cunha:2019dwb}
	\begin{eqnarray}
		S&=& \frac{1}{2\kappa}\int dx^4\sqrt{-g}\left(R-2\Lambda +\alpha{\cal L}_{GB}+{\cal L}_{\Phi}\right),\nonumber\\
		{\cal L}_{GB}&=& f(\Phi){\cal G},\nonumber\\
		{\cal L}_\Phi&=&-\frac{1}{2}\nabla^\mu\Phi\nabla_\mu\Phi-V(\Phi),\nonumber\\
		{\cal G}&=&R^2+R_{\mu\nu\rho\sigma}R^{\mu\nu\rho\sigma}-4R_{\mu\nu}R^{\mu\nu},
	\end{eqnarray}
	where the scalar field $\Phi$ is non-minimally coupled to the Gauss-Bonnet term ${\cal G}$ with the coupling constant $\alpha$. $f(\Phi)$ is a function of the scalar field and $\Lambda$ is the cosmological constant. From the action, one can derive the equations of motion
	\begin{eqnarray}
		\nabla^2\Phi&=&\frac{dV}{d\Phi}-\alpha \frac{d f}{d\Phi}{\cal G},\label{ScalarEq}\\
		R_{\mu\nu}-\frac{1}{2}g_{\mu\nu}R + \Lambda g_{\mu\nu}&=&\alpha T^{GB}_{\mu\nu}+T^\Phi_{\mu\nu},\label{MetricEq}\\
		T^{GB}_{\mu\nu}&=&2(\nabla_\mu\nabla_\nu f)R-2g_{\mu\nu}(\nabla_\rho\nabla^\rho f)R \nonumber\\
		&&-4(\nabla^\rho\nabla_\nu f)R_{\mu\rho}-4(\nabla^\rho\nabla_\mu f)R_{\nu\rho}\nonumber\\
		&&+4(\nabla^\rho\nabla_\rho f)R_{\mu\nu}+4g_{\mu\nu}(\nabla^\rho\nabla^\sigma f)R_{\rho\sigma}\nonumber\\
		&&-4(\nabla^\rho\nabla^\sigma f)R_{\mu\rho\nu\sigma},\nonumber\\
		T^\Phi_{\mu\nu}&=&\frac{1}{2}\nabla_\mu\Phi\nabla_\nu\Phi-\frac{1}{2} g_{\mu\nu}V(\Phi)-\frac{1}{4} g_{\mu\nu}\nabla^\rho\Phi\nabla_\rho \Phi.\nonumber
	\end{eqnarray}
	The theory admits GR black hole solutions with constant scalar profile $\Phi=\Phi_0$,  if
	\begin{eqnarray}
		V(\Phi_0)=0,\quad \frac{dV}{d\Phi}\bigg|_{\Phi_0}=0, \quad \frac{df}{d\Phi}\bigg|_{\Phi_0}=0.
	\end{eqnarray}
	In the following, we will consider a simple case as  \cite{Cunha:2019dwb} by choosing $\Lambda=0, V(\Phi)=0$ and
	\begin{eqnarray}
		f(\Phi)=\frac{1}{2\beta} \left(1-e^{-\beta \Phi^2}\right),
	\end{eqnarray}
	where $\beta>0$ is a constant. 
	
	We are going to study the wave dynamics of scalar field perturbations on the background of GR black holes in the linear regime. More precisely, we consider  the Kerr black hole with $\Phi=0$. The metric in the
	Boyer-Lindquist coordinates is
	\begin{eqnarray}
		ds^2 = -\frac{\Delta}{\Sigma} (dt - a \sin^2 \theta d\phi)^2 + \frac{\Sigma}{\Delta} dr^2 + \Sigma d\theta^2 + \frac{\sin^2\theta}{\Sigma} (a dt - (r^2+a^2)d\varphi)^2,
	\end{eqnarray}
	where $\Delta\equiv r^2 - 2Mr+a^2$ and $\Sigma \equiv r^2 + a^2 \cos^2\theta$. In this case, the scalar perturbation equation (\ref{ScalarEq}) in the Kerr background reduces to
	\begin{eqnarray}
		&&\nabla^2\Phi=-\alpha {\cal G} \Phi, \label{ScalarEq2}\\
		&&{\cal G} = -\frac{48 M^2 \left(-32 r^6+48 r^4 \Sigma -18 r^2 \Sigma^2+\Sigma^3\right)}{\Sigma^{6}},\nonumber
	\end{eqnarray}
	where ${\cal G}$ is valued in the background. As one can see,  the scalar field  has an effective mass $m^2_{\rm eff}= -\alpha {\cal G}$, which is position dependent and approaches zero at infinity, with the sign depending on the coupling constant $\alpha$.
	When $\alpha=0$, the above equation describes wave propagations of a free scalar field in the Kerr background which has been studied thoroughly  \cite{Krivan:1996da,PazosAvalos:2004rp,Thuestad:2017ngu}. For  $\alpha>0$,  in the Schwarzschild background, one has ${\cal G} = 48 M^2/r^6$ so that the effective mass square is always negative, which leads to the tachyonic instability and the subsequent spontaneous scalarization. In \cite{Doneva:2017bvd,Silva:2017uqg} it was found that the instability occurs when $M/(2 \sqrt{\alpha}) \lesssim 0.587$ (note that in \cite{Doneva:2017bvd,Silva:2017uqg}  $\lambda^2$ was used to represent the coupling constant which is related to $\alpha$ by $\lambda^2 = 4 \alpha$), and becomes more violent for larger $\alpha$ or smaller $M$. However, for a Kerr background, the situation becomes more complicated, because  ${\cal G}$ is not a  monotonic function so that the effective mass square is not always negative. How does this effective mass affects the stability of the Kerr black hole configuration and triggers spontaneous scalarization  requires careful investigations. 
	
	In the following sections, we will study carefully the time evolution of the scalar perturbation and obtain  object pictures on the influences of the coupling constant $\alpha$ on wave dynamics in the sEGB theory.

	\section{Setup of the numerical method}
	
	As mentioned in the introduction, traditional numerical methods usually encounter the so-called ``outer boundary problem", which can spoil the numerical precision, especially for the evolution of the scalar perturbation at the late time \cite{Thuestad:2017ngu}. To overcome this problem, we  apply the numerical strategy proposed in \cite{Zenginoglu:2007jw,Zenginoglu:2008pw,Zenginoglu:2008uc,Zenginoglu:2008wc,Zenginoglu:2009ey,Zenginoglu:2009hd,Zenginoglu:2010cq,Zenginoglu:2011jz,Zenginoglu:2011zz} to solve the scalar perturbation equation (\ref{ScalarEq2}), which contains two steps of coordinates transformations mainly. First, we define the ingoing coordinates $\{\tilde{t}, r, \theta, \tilde{\varphi}\}$ through the following transformation
	\begin{eqnarray}
		d\tilde{t} = dt + \frac{2 M r}{\Delta} dr,\qquad d\tilde{\varphi} = d\varphi + \frac{a}{\Delta} dr,
	\end{eqnarray}
	and the metric becomes
	\begin{eqnarray}
		ds^2 =&& -\left(1-\frac{2 M r}{\Sigma}\right) d\tilde{t}^2 - \frac{4 a M r}{\Sigma} \sin^2\theta d\tilde{t} d\tilde{\varphi} + \frac{4 M r}{\Sigma} d\tilde{t} dr\nonumber\\
		&& + \left(1+\frac{2 M r}{\Sigma}\right) dr^2 - 2 a \sin^2\theta \left(1+\frac{2 M r}{\Sigma}\right) dr d\tilde{\varphi} + \Sigma d\theta^2\nonumber\\
		&& + \left(r^2 + a^2 + \frac{2 M a^2 r \sin^2\theta}{\Sigma}\right) \sin^2\theta d\tilde{\varphi}. \label{KerrSchildMetric}
	\end{eqnarray}
	Considering the axial symmetry of the Kerr spacetime, the scalar perturbation can be decomposed as
	\begin{eqnarray}
		\Phi (\tilde{t},r,\theta,\tilde{\varphi}) = \frac{1}{r} \sum_m \Psi(\tilde{t},r,\theta) e^{i m \tilde{\varphi}}.
	\end{eqnarray}
	Substituting the above ansatz into  Eq.~(\ref{ScalarEq2}), the scalar perturbation equation in the ingoing  coordinates becomes
	\begin{eqnarray}
		A^{\tilde{t}\tilde{t}} \partial_{\tilde{t}}^2 \Psi + A^{\tilde{t} r} \partial_{\tilde{t}} \partial_{r} \Psi + A^{r r} \partial_{r}^2 \Psi + A^{\theta\theta} \partial_{\theta}^2 \Psi + B^{\tilde{t}} \partial_{\tilde{t}} \Psi + B^r \partial_r \Psi + B^\theta \partial_{\theta}\theta \Psi + C \Psi =0,
	\end{eqnarray}
	where 
	\begin{eqnarray}
		A^{\tilde{t}\tilde{t}} &=& \Sigma + 2 M r,\nonumber\\
		A^{\tilde{t} r} &=& -4 M r,\nonumber\\
		A^{r r} &=& -\Delta,\nonumber\\
		A^{\theta\theta} &=& -1,\nonumber\\
		B^{\tilde{t}} &=& 2 M,\nonumber\\
		B^r &=& \frac{2}{r} (a^2 - M r) - 2 i m a,\nonumber\\
		B^\theta &=& -\cot\theta,\nonumber\\
		C &=& \frac{m^2}{\sin^2\theta}- \frac{2(a^2-M r)}{r^2} + \frac{2 i m a}{r} - \alpha \Sigma \cal{G}.
	\end{eqnarray}
	
	The second step is to use the technique of hyperboloidal compactifications developed in \cite{Harms:2014dqa},  which defines a compactified horizon-penetrating, hyperboloidal coordinates $(\tau,\rho,\theta,\tilde{\varphi})$ (HH coordinates) through the transformation \footnote{Of course, in this step, one can have other choices, such as the RT coordinates proposed in  \cite{Racz:2011qu}.}
	\begin{eqnarray}
		\tilde{t} = \tau + h(\rho),\qquad r = \frac{\rho}{\Omega(\rho)},\label{HHCoordinates}
	\end{eqnarray}
	where 
	\begin{eqnarray}
		h(\rho) = \frac{\rho}{\Omega} - \rho -4 M \ln\Omega,\qquad \Omega (\rho) = 1- \frac{\rho}{S}.
	\end{eqnarray}
	Here $S$ is a free parameter controlling both the domain and the foliation. With this compactification, the semi-infinite radial domain outside the horizon $r\in [r_+, \infty)$ is mapped to a finite range $\rho \in [\rho_+, S)$ with the event horizon $r=r_+$ now locates at 
	\begin{eqnarray}
		\rho_+ = \frac{a^2 S + M S^2 + \sqrt{M^2 S^4 - a^2 S^4}}{a^2 + 2 M S +S^2}.
	\end{eqnarray}
	Then we have the relations
	\begin{eqnarray}
		\partial_{\tilde{t}} = \partial_\tau,\qquad \partial_r = - H \partial_\tau + K \partial_\rho,
	\end{eqnarray}
	where $H \equiv \frac{d h}{dr} (\rho)$ and $K \equiv \frac{d\rho}{dr} (\rho)$. With these relations, the scalar perturbation equation in the HH coordinates can be written as
	\begin{eqnarray}
		\partial_\tau^2 \Psi = \tilde{A}^{\tau\rho} \partial_\tau \partial_\rho \Psi + \tilde{A}^{\rho\rho} \partial_\rho^2 \Psi + \tilde{A}^{\theta\theta} \partial_{\theta}^2 \Psi + \tilde{B}^\tau \partial_\tau \Psi + \tilde{B}^\rho \partial_\rho \Psi + \tilde{B}^\theta \partial_{\theta} \Psi + \tilde{C} \Psi,  
	\end{eqnarray}
	where 
	\begin{eqnarray}
		\{\tilde{A}^{\tau\rho}, \tilde{A}^{\rho\rho}, \tilde{A}^{\theta\theta}, \tilde{A}^\tau, \tilde{B}^\tau, \tilde{B}^\rho, \tilde{B}^\theta, \tilde{C}\} = -\frac{1}{A^{\tau\tau}} \{A^{\tau\rho}, A^{\rho\rho}, A^{\theta\theta}, A^\tau, B^\tau, B^\rho, B^\theta, C\},
	\end{eqnarray}
	and 
	\begin{eqnarray}
		A^{\tau\tau} &=& A^{\tilde{t}\tilde{t}} - H A^{\tilde{t} r} + H^2 A^{rr},\nonumber\\
		A^{\tau\rho} &=& K A^{\tilde{t} r} - 2 K H A^{rr},\nonumber\\
		A^{\rho\rho} &=& K^2 A^{rr},\nonumber\\
		B^\tau &=& B^{\tilde{t}} - H B^r - \frac{d H}{d\rho} K A^{rr},\nonumber\\
		B^\rho &=& K \left(B^r + \frac{d K}{d\rho} A^{rr}\right),
	\end{eqnarray}
	To solve the above equation numerically in the time domain, it is convenient to rewrite it as two coupled first-order partial differential equations by introducing a new variable $\Pi \equiv \partial_\tau \Psi$,
	\begin{eqnarray}
		\partial_\tau \Psi &=& \Pi,\nonumber\\
		\partial_\tau \Pi &=& \tilde{B}^\tau \Pi + \tilde{A}^{\tau\rho} \partial_\rho \Pi + \tilde{A}^{\rho\rho} \partial_\rho^2 \Psi + \tilde{B}^\rho \partial_\rho \Psi+ \tilde{A}^{\theta\theta} \partial_{\theta}^2 \Psi   + \tilde{B}^\theta \partial_{\theta} \Psi + \tilde{C} \Psi.
	\end{eqnarray} 
	
	Numerically, this form is suitable for the method mentioned above \cite{Schiesser}. Precisely, the derivatives in $\rho$ and $\theta$ directions are approximated by finite differences, and the evolution in
	the time direction is implemented with the fourth-order Runge-Kutta integrator. One of the main advantages of working in the HH coordinates is that the ingoing (outgoing) boundary condition 
	at the horizon (infinity) is satisfied automatically, so we do not need to worry about the sophisticated ``outer boundary problem" which is hard to handle and influences the precision. At the poles in 
	the angular direction $\theta=0$ and $\pi$, physical boundary conditions are needed
	\begin{eqnarray}
		\Psi|_{\theta=0, \pi} &=& 0 \quad {\rm for \ odd} \ m,\\
		\partial_\theta \Psi|_{\theta=0, \pi} &=& 0 \quad {\rm for \ even}\  m.
	\end{eqnarray}
	Following \cite{PazosAvalos:2004rp}, we use a staggered grid and add ghost points to implement these conditions.

	\section{Results}
	
	We consider the initial data of the scalar perturbation to be a Gaussian distribution localized outside the horizon at $\rho=\rho_c$,
	\begin{eqnarray}
		\Psi (\tau=0,\rho,\theta) \sim Y_{\ell m} e^{-\frac{(\rho-\rho_c)^2}{2 \sigma^2}},
	\end{eqnarray}
	where $Y_{\ell m}$ is the $\theta$-dependent part of the spherical harmonic function and $\sigma$ is the width of the Guassian distribution. Without loss of generality, we consider the perturbation to have time symmetry so that
	\begin{eqnarray}
		\Pi (\tau=0,\rho,\theta)=0.
	\end{eqnarray}
	In the following, we take $\rho_c=6 M$. Also, we set $M=1$ so that all quantities are measured in units of $M$. Observers are assumed to locate at $\rho =6 M$ and $\theta=\frac{\pi}{4}$.  We choose the free parameter $S=10$ as in \cite{Harms:2014dqa}. 
	
	The Kerr spacetime is not spherically symmetric, except when $a=0$, so the mode-mixing phenomenon occurs \cite{Zenginoglu:2012us,Burko:2013bra,Thuestad:2017ngu}: a pure even (odd) initial $\ell$ multipole will excite other  even (odd) multipoles (denoted by $\ell'$) with the same $m$ as it  evolves. So, in this work we only consider $\ell=0$ and $\ell=1$ as representative even and odd multipoles, respectively. Also, we only consider axisymmetric perturbations with $m=0$ for simplicity.
	
	\subsection{$\alpha=0$}
	
	To check the reliability of our numerical strategy so as to further disclose  precisely the influence of $\alpha$ on the wave dynamics, we first consider the simple case where the scalar field does not couple to the GB term, i.e., $\alpha=0$. We compare our results with those in the ringdown phase and the late-time tail with available results either in the time domain or in the frequency domain. In this case with zero $\alpha$, the scalar field equation is separable for the $r$ and $\theta$ variables and can be cast into a Schr$\ddot{o}$dinger-like form. Then, QNMs can be computed directly in the frequency domain by using Leaver's continued fraction method \cite{Kokkotas:1999bd,Nollert:1999ji,Berti:2009kk,Konoplya:2011qq}.

	\begin{figure}[!htbp]
		\centering
		\subfigure[$~~a=0$ ]{\includegraphics[width=0.47\textwidth]{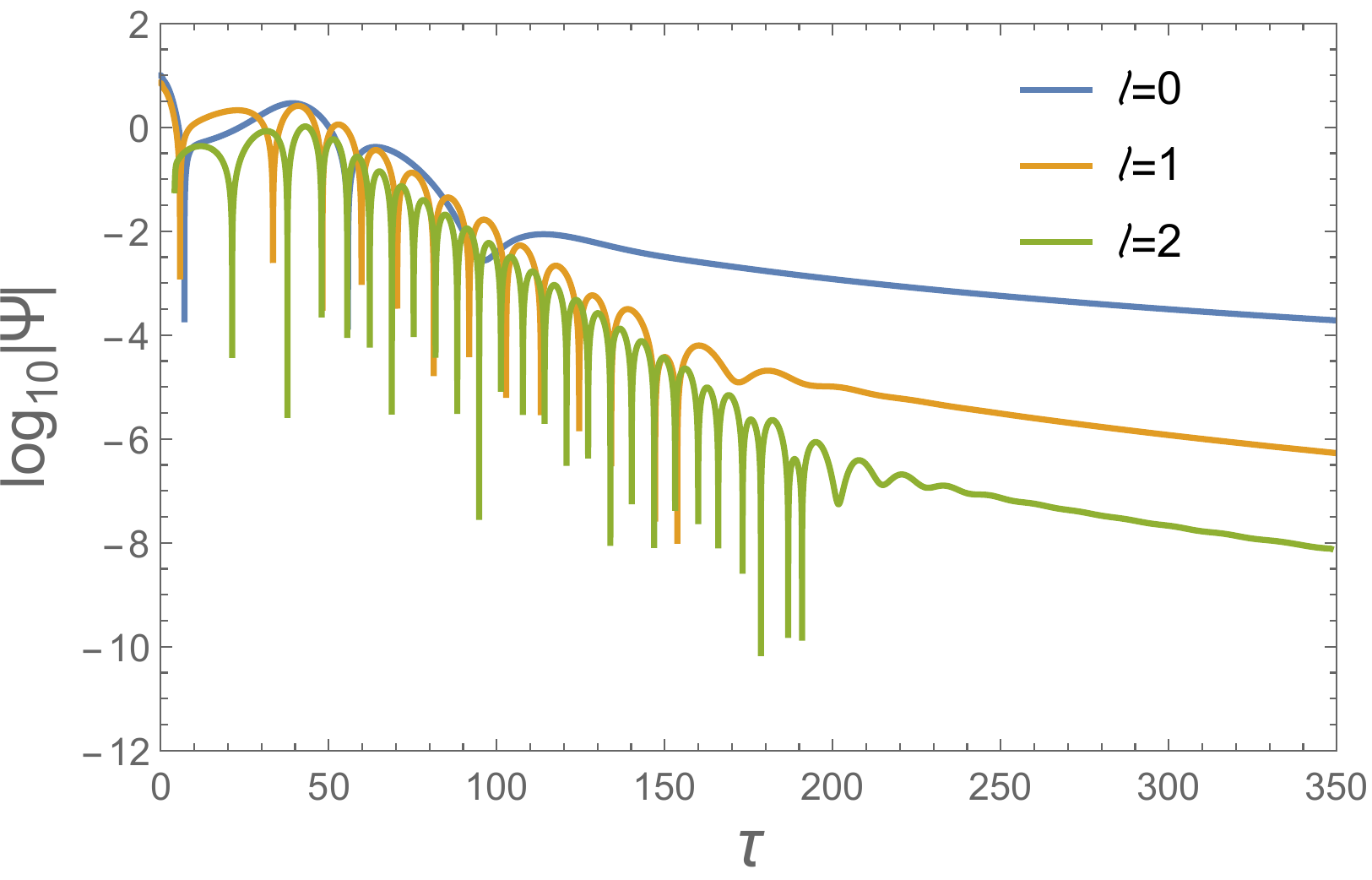}} \quad
		\subfigure[$~~a=0.5$ ]{\includegraphics[width=0.47\textwidth]{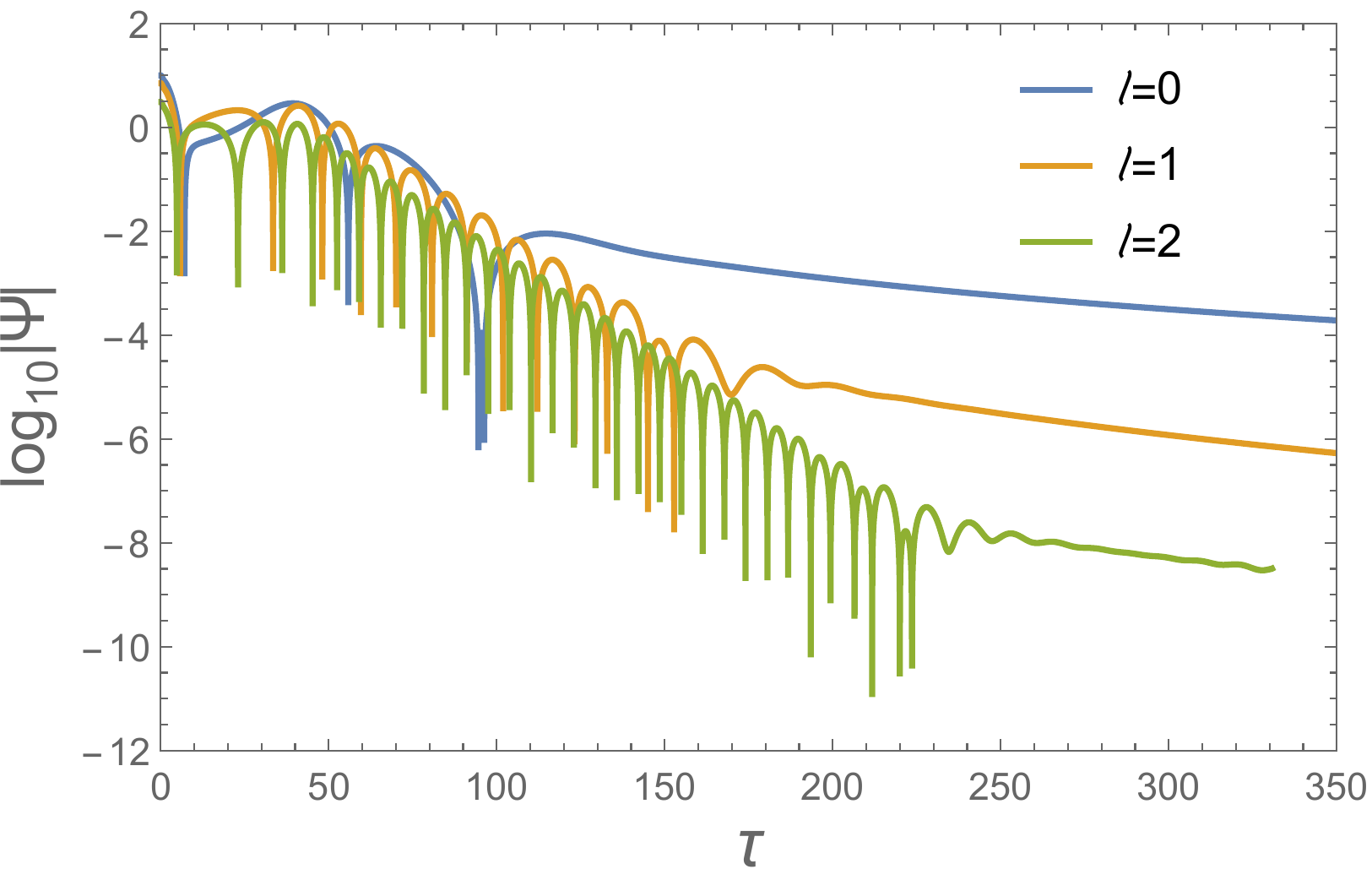}}
		\subfigure[$~~a=0.9$ ]{\includegraphics[width=0.47\textwidth]{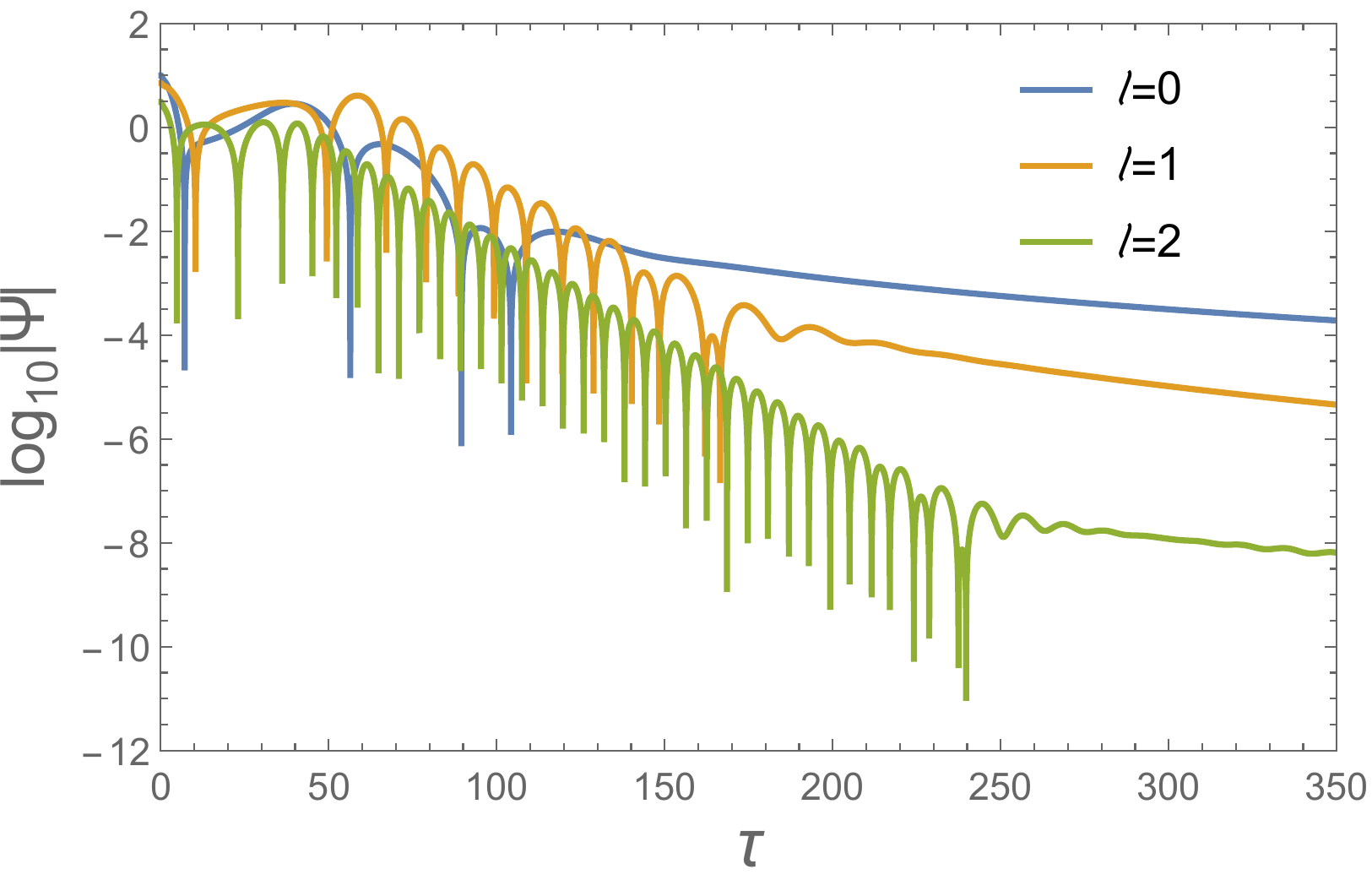}}
		\caption{(color online) Time evolution of the axisymmetric scalar perturbation for $a=0, 0.5$ and $0.9$. The initial multipoles we considered are $\ell=0, 1$ and $2$. Time is in units of $M$.}
		\label{alpha0Figs}
	\end{figure}
	
	In Fig. \ref{alpha0Figs}, the time evolutions of the scalar perturbation in the Kerr background are plotted for various spins. We only consider axisymmetric initial multipoles ($m=0$) with various $\ell$'s. From the figure, we can read off the quasinormal frequencies by using the Prony's method \cite{Berti:2007dg}. The results are listed in Table \ref{QNMsIndex}  denoted as $\omega_{\rm Prony}$. To make a comparison, the results derived by Leaver's method in the frequency domain \cite{Berti:2009kk,Dolan:2007mj} are also listed denoted as $\omega_{\rm Leaver}$. From the table, we can see that using the numerical method described in the last section, our obtained results  agree well with previous ones. The relatively large discrepancy for the $\ell=0$ case is due to the short duration of the ringdown phase so that Prony's method can not give quasinormal frequencies with sufficient precisions.
	
	\begin{table}[!htbp]
		\centering
		\begin{tabular}{p{1cm}<{\centering} p{1cm}<{\centering} p{3cm}<{\centering} p{3cm}<{\centering} p{1cm}<{\centering} p{2cm}<{\centering}}
			\hline
			\hline
			$a$ & $\ell$ & $\omega_{\rm Prony}$ & $\omega_{\rm Leaver}$ & $n_{\rm fitting}$ & $n_{\rm predicted}$\\
			\hline
			\multirow{3}*{$0$} & $0$ & $0.1084-0.1062i$ & $0.1105-0.1049i$ & $3.16$ & $3$\\
			&$1$ & $0.2929-0.0976i$ & $0.2929-0.0977i$ & $5.10$ & $5$\\
			&$2$ & $0.4836-0.0968i$ & $0.4836-0.0968i$ & $7.32$ & $7$\\
			\hline
			\multirow{3}*{$0.5$} & $0$ & $0.1124-0.0935i$ & $0.1123-0.1022i$ & $3.16$ & $3$\\
			& $1$ & $0.2979-0.0954i$ & $0.2979-0.0954i$ & $5.18$ & $5$\\
			& $2$ & $0.4920-0.0946i$ & $0.4920-0.0946i$ & $3.38$ & $3$\\ 
			\hline
			\multirow{3}*{$0.9$} & $0$ & $0.1186-0.0901i$ & $0.1138-0.0916i$ & $3.18$ & $3$\\
			& $1$ & $0.3108-0.0866i$ & $0.3108-0.0866i$ & $5.18$ & $5$\\
			& $2$ & $0.5148-0.0864i$ & $0.5148-0.0864i$ & $3.35$ & $3$\\ 
			\hline\hline
		\end{tabular}
		\caption{Dominant QNMs and the power-law index of the late-time tail for axisymmetric scalar perturbations with $a=0, 0.5$ and $0.9$. $n_{\rm predicted}$ refers to the power-law index according to Eq. (\ref{PowerLaw}) and $n_{\rm fitting}$ is our fitting result.}
		\label{QNMsIndex}
	\end{table}	
	
	Now let us examine the late-time tail. Interestingly, it is only recently and with the hyperboloidal compactification technique that we have a precise and well understanding of the late-time behaviors of the scalar field perturbation in the Kerr black hole background \cite{Zenginoglu:2012us,Burko:2013bra,Thuestad:2017ngu}. For axisymmetric perturbations ($m=0$), the late-time behaviors of these multipoles exhibit a power-law decay $\Psi \sim \tau^{-n}$ at time infinities, where
	\begin{eqnarray}
		n = \left\{
		\begin{array}{ll}
			\ell+\ell'+3 \quad {\rm for} \ \ell=0,1\\
			\ell+\ell'+1 \quad {\rm otherwise}
		\end{array}
		\right. \label{PowerLaw}
	\end{eqnarray}
	In the full late-time, the dominant mode is $\ell'=0$ for the even $\ell$ case and $\ell'=1$ for the odd $\ell$  case (for more details we refer readers to Refs. \cite{Zenginoglu:2012us,Burko:2013bra,Thuestad:2017ngu}).

	From Fig. \ref{alpha0Figs}, we can see that in the late-time, the perturbation exhibits a power-law decay as expected. By fitting the late-time data with function $\tau^{-n}$, we can find the index that is listed in Table \ref{QNMsIndex} denoted as $n_{\rm fitting}$. It should be noted that the power-law decay (\ref{PowerLaw}) only holds exactly at very late-time, so to make a precise comparison we need to evolve the perturbation to a very late time which will be extremely time consuming. However, in the time range we show that the index $n_{\rm fitting}$ and its trend with time derived by fitting agrees well with the prediction (\ref{PowerLaw}).
	
	\subsection{$\alpha>0$}
	
\begin{figure}[!htbp]
		\centering
		\subfigure[$~~a=0$ ]{\includegraphics[width=0.47\textwidth]{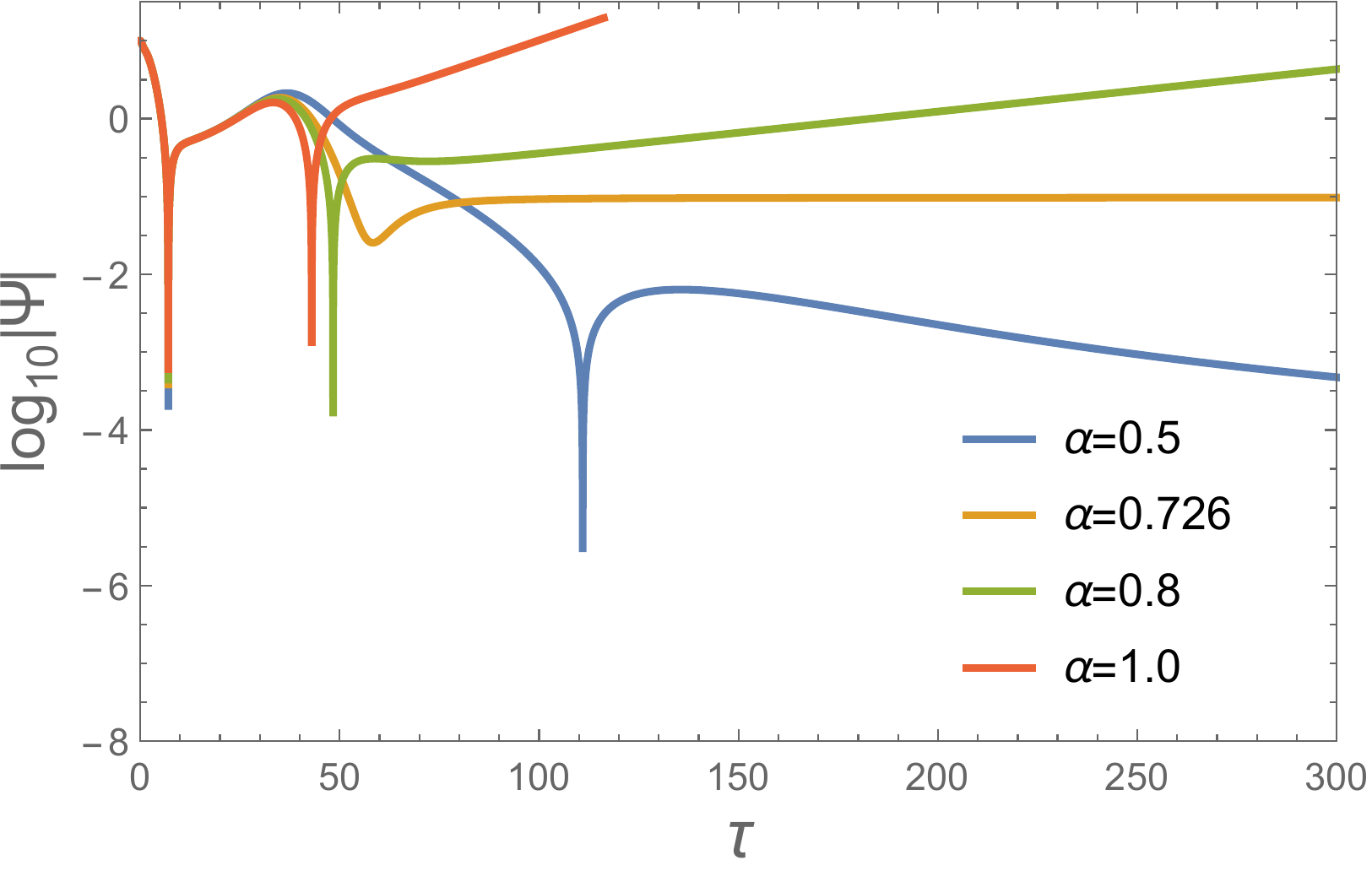}} \quad
		\subfigure[$~~a=0.3$ ]{\includegraphics[width=0.47\textwidth]{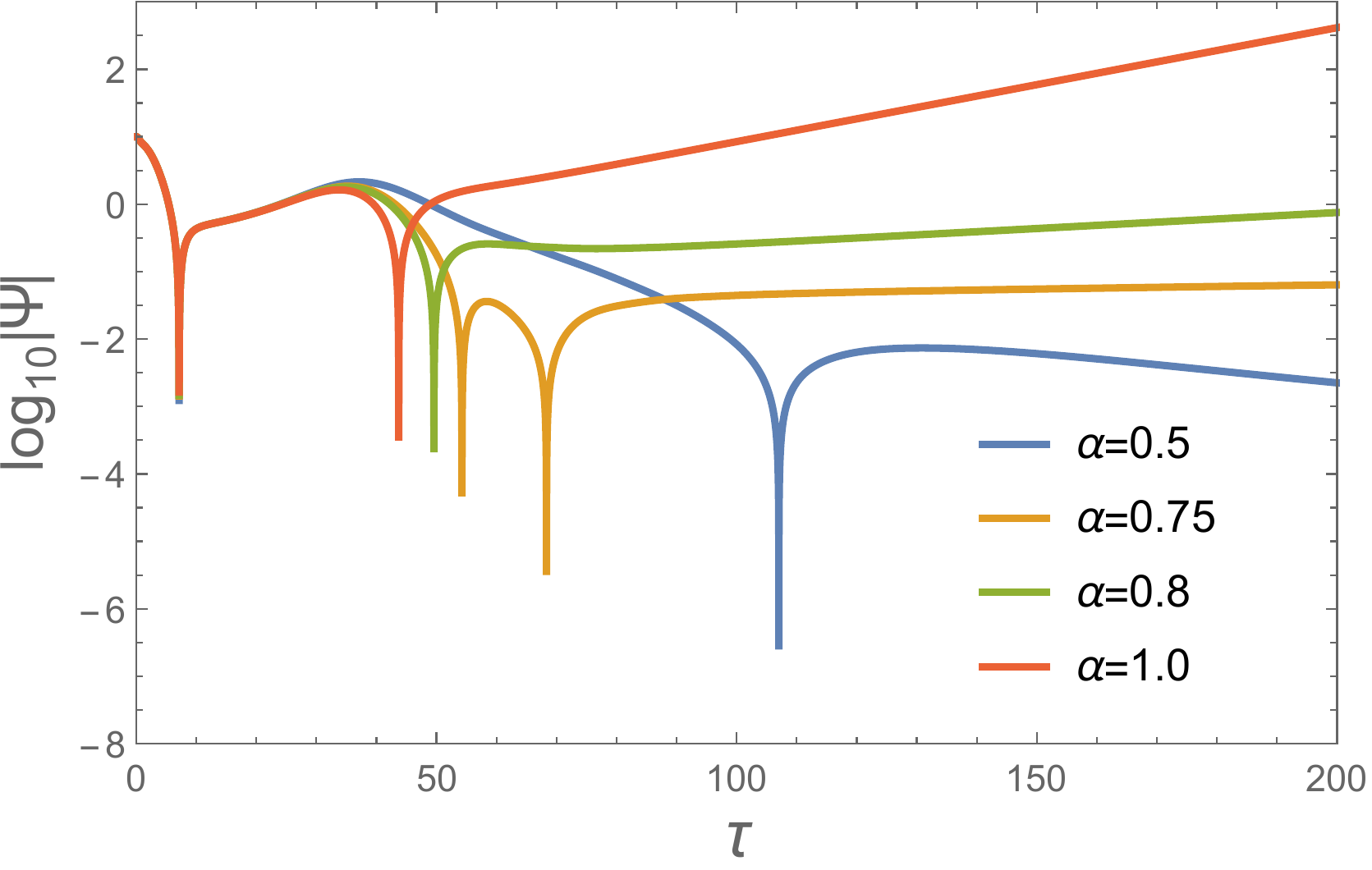}}
		\subfigure[$~~a=0.5$ ]{\includegraphics[width=0.47\textwidth]{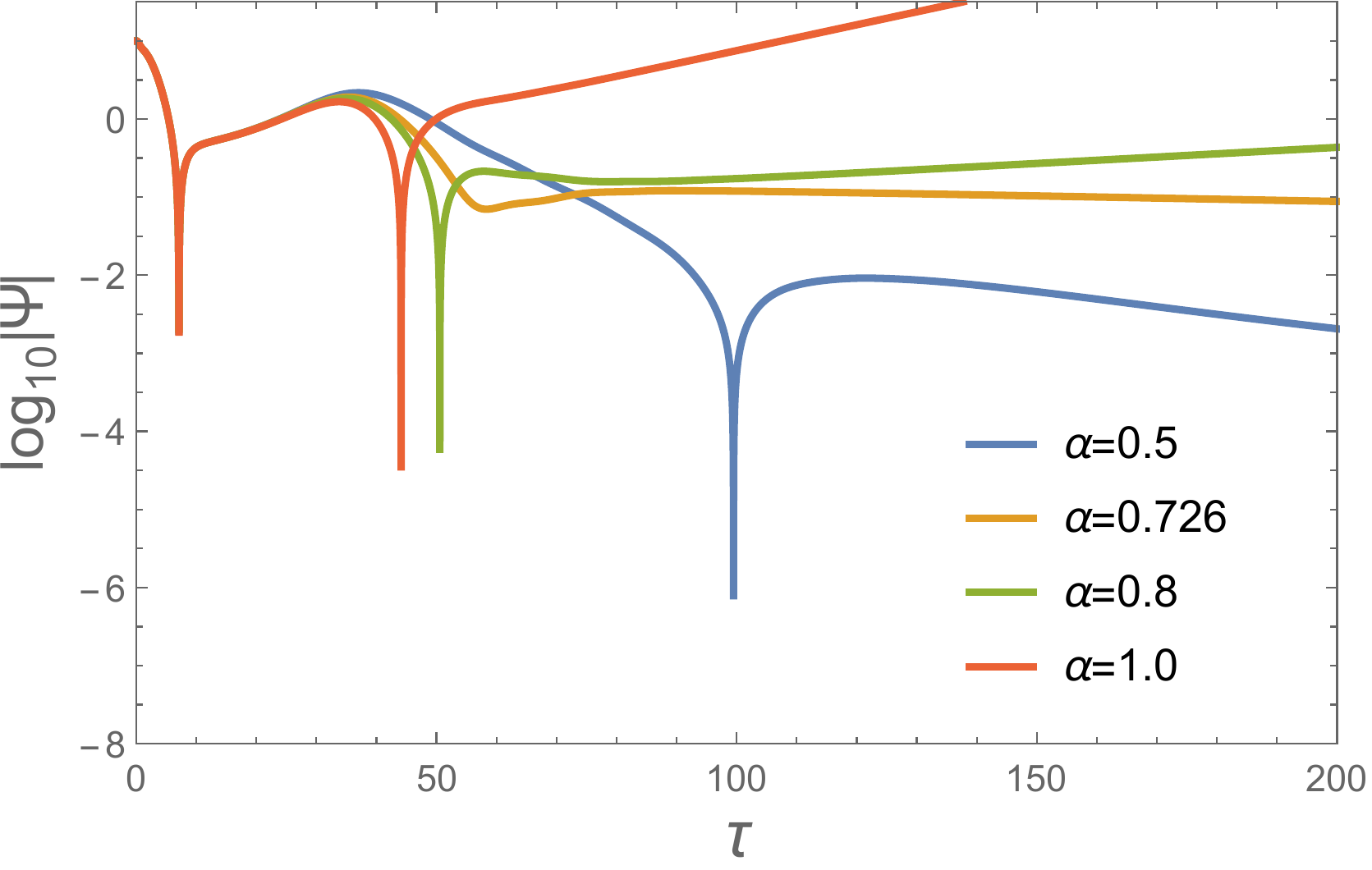}}\quad
		\subfigure[$~~a=0.9$ ]{\includegraphics[width=0.47\textwidth]{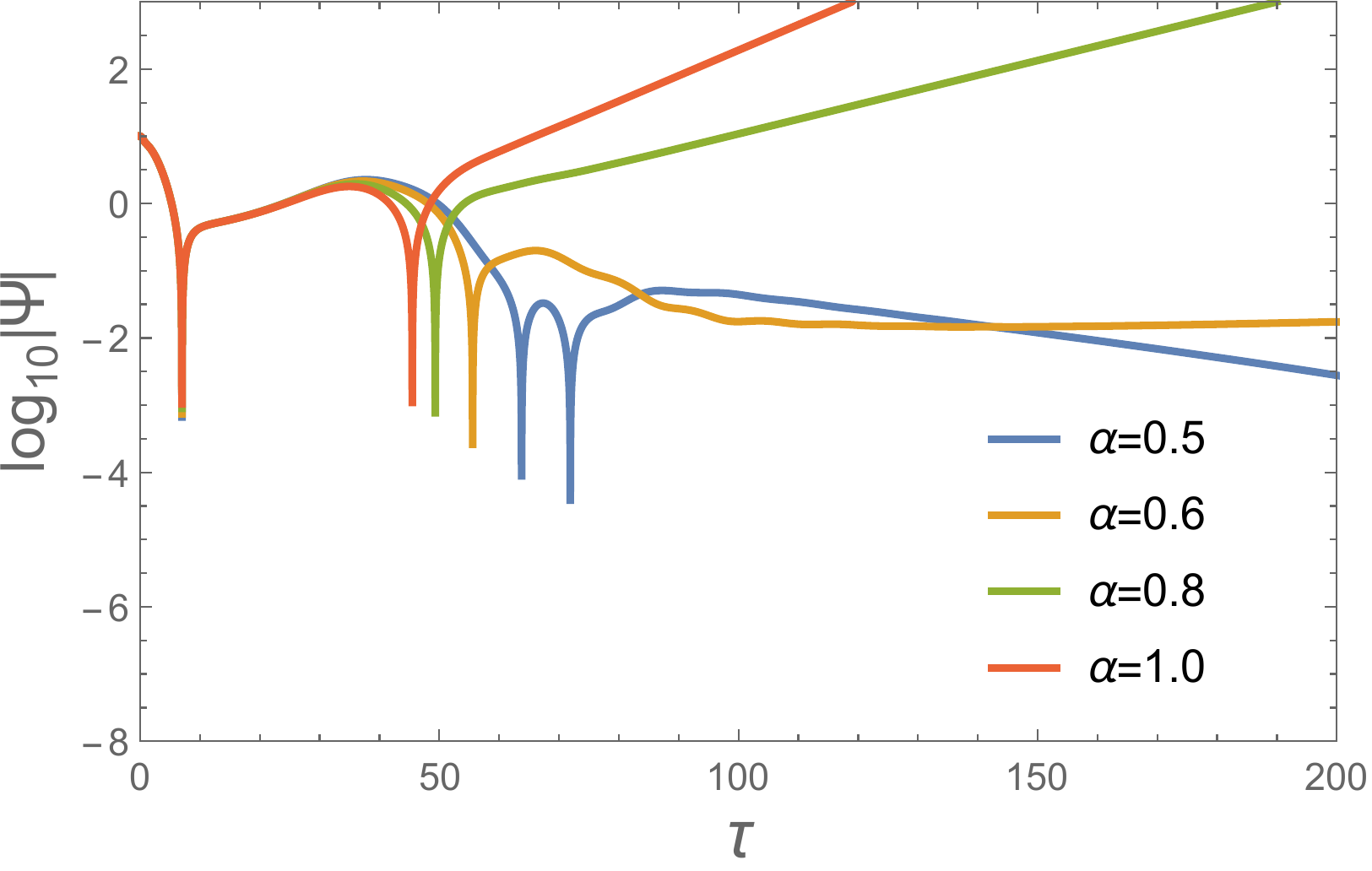}}
		\caption{(color online) Time evolution of the scalar perturbation for $a=0, 0.3, 0.5$ and $0.9$ with $\alpha>0$.  Initial multipole is fixed as $\ell=0$. Time is in units of $M$.}
		\label{alphaPositiveL0m0Fig}
	\end{figure}

	Now let us consider the case with $\alpha>0$. In Fig. \ref{alphaPositiveL0m0Fig}, the time evolution behaviors of the scalar perturbation for various positive $\alpha$'s are plotted. The initial multipole is taken as $\ell=0$. We first look at the wave propagation after the ringing. From the figure, one can see that when $a=0$, there exists a critical value  $\alpha_c \sim 0.726$, above which the perturbation grows,  indicating the occurrence of the instability. The instability happens earlier and becomes more violent as $\alpha$ further increases. Similar behaviors of the stability also appear for $a \neq 0$.  We learn that the instability starts at $\alpha_c=  0.75, 0.726, 0.71$ and $0.6$  for different rotations $a=0.3, 0.5, 0.7$ and $0.9$,  respectively. The dependence of $\alpha_c$ on $a$  is not simply monotonic as the increase of the spin from zero to the extreme value, $\alpha_c$ first increases when the hole starts to rotate and then changes to decrease as the hole rotates faster until the extreme value. Using a different and more precise numerical strategy described in the last section, we have shown  that our results agree quantitatively well with those computed in  \cite{Doneva:2017bvd,Silva:2017uqg}. 
	
	\begin{figure}[!htbp]
		\centering
		\subfigure[$~~a=0$ ]{\includegraphics[width=0.47\textwidth]{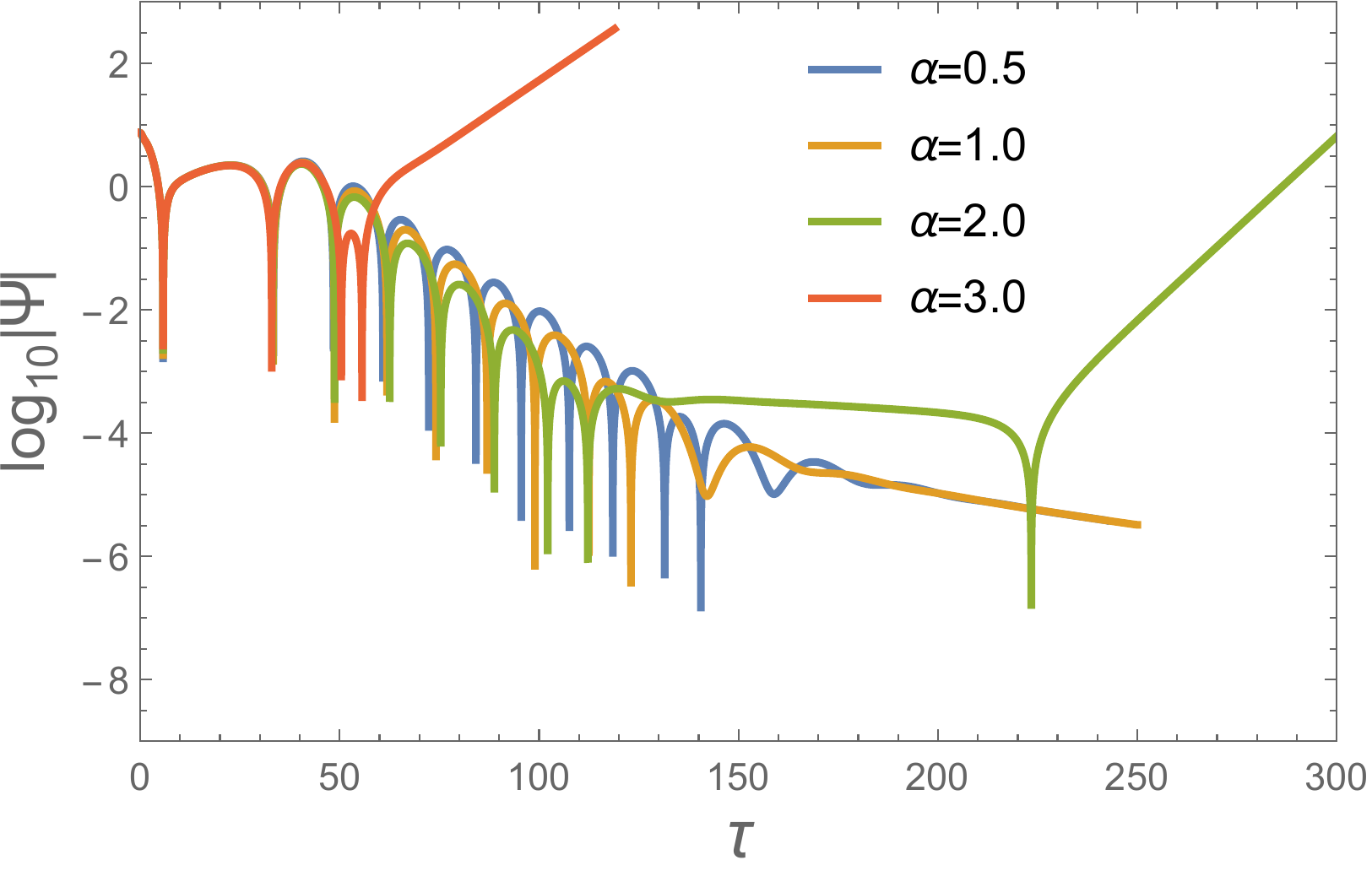}} \quad
		\subfigure[$~~a=0.5$ ]{\includegraphics[width=0.47\textwidth]{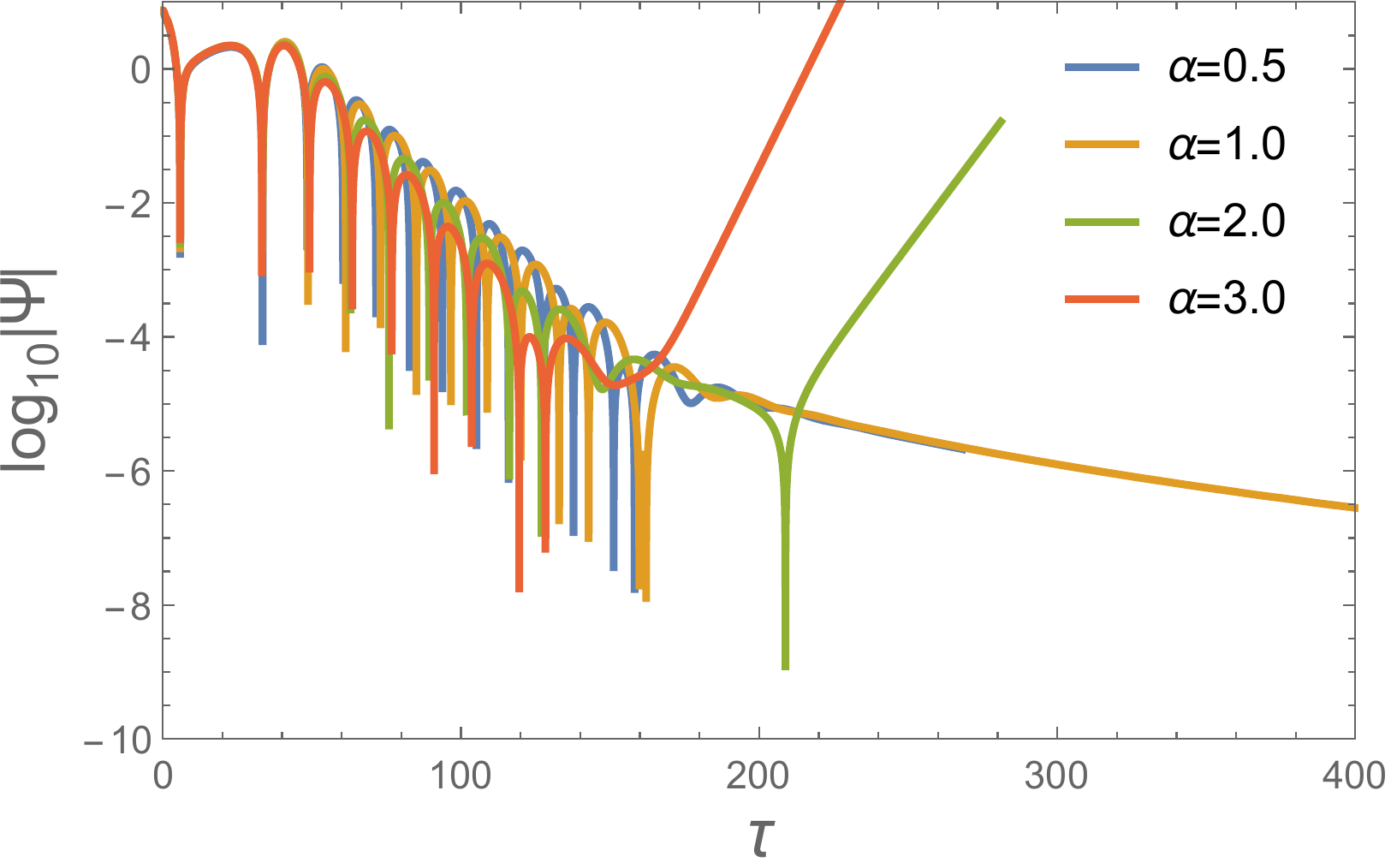}}
		\subfigure[$~~a=0.9$ ]{\includegraphics[width=0.47\textwidth]{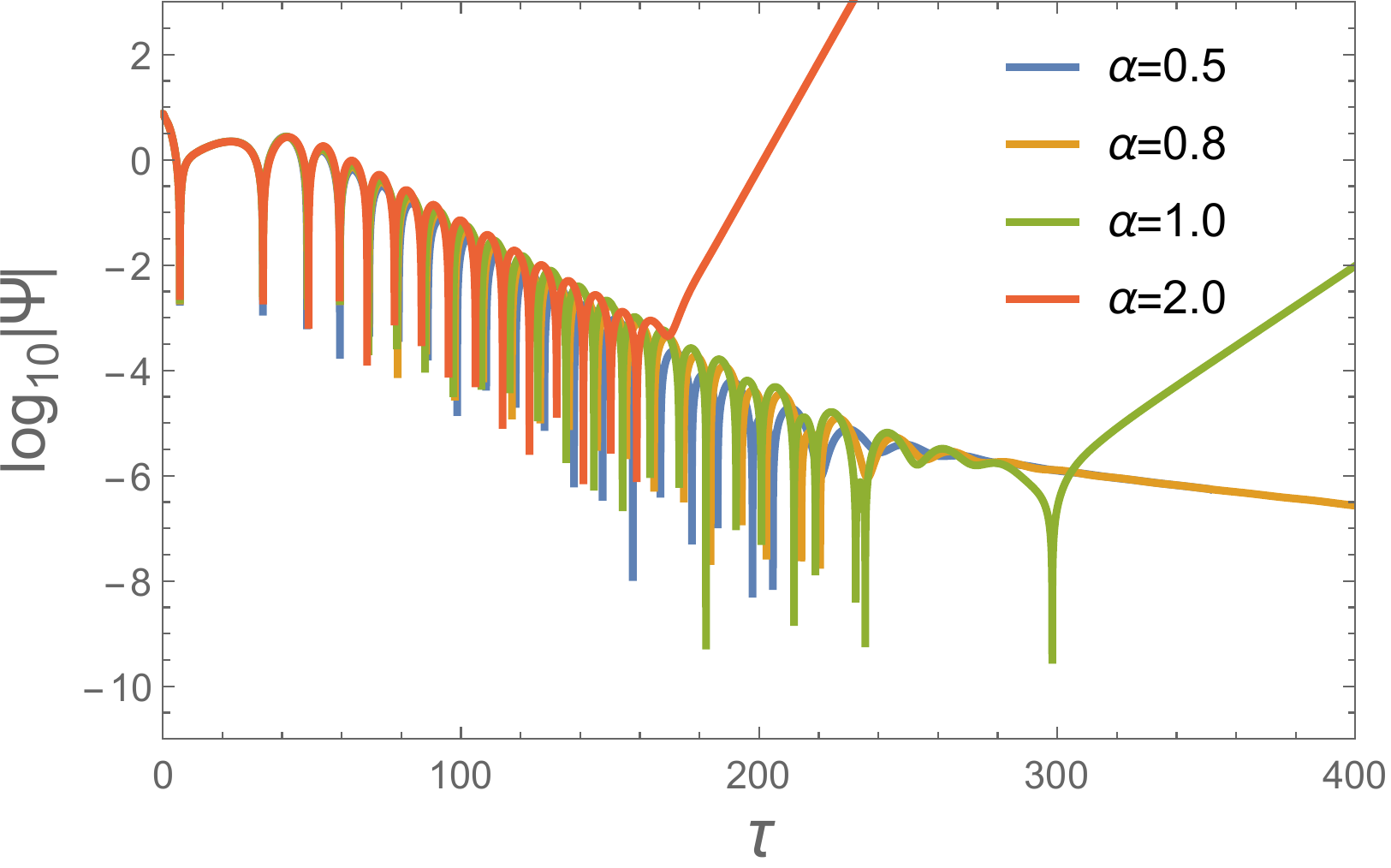}}
		\caption{(color online) Time evolution of the scalar perturbation for $a=0, 0.5$ and $0.9$ with $\alpha>0$. Initial multipole is fixed as $\ell=1$. Time is in units of $M$.}
		\label{alphaPositiveL1m0Fig}
	\end{figure}
	
	\begin{table}[!htbp]
		\centering
		\begin{tabular}{p{1cm}<{\centering} p{3cm}<{\centering} p{3cm}<{\centering} p{3cm}<{\centering}}
			\hline
			\hline
			\multirow{2}*{$\alpha$} & \multicolumn{3}{c}{$a$}\\
			\cline{2-4}
			&  $0$ & $0.5$ & $0.9$ \\
			\hline
			$-2.0$ & $0.3870-0.1105i$  & $0.3613-0.1067i$  & $-$\\
			$-1.5$ & $0.3633-0.1062i$  & $0.3456-0.1034i$  & $0.3387-0.1150i$\\
			$-1.0$ & $0.3397-0.1020i$ & $0.3300-0.1001i$  & $0.3118-0.1122i$\\
			$-0.5$ & $0.3163-0.0988i$ & $0.3139-0.0975i$ & $0.3088-0.1019i$ \\
			$0$ & $0.2929-0.0976i$ &  $0.2979-0.0954i$ & $0.3108-0.0866i$\\
			$0.5$ & $0.2703-0.1002i$ & $0.2817-0.0945i$ &  $0.3210-0.0747i$\\
			$0.8$ & $0.2580-0.1046i$ &  $0.2721-0.0948i$ & $0.3281-0.0711i$\\
			$1.0$ & $0.2512-0.1088i$ &  $0.2657-0.0955i$ & $0.3325-0.0700i$\\
			$2.0$ & $0.2394-0.1244i$ &  $0.2389-0.1067i$ & $0.3467-0.0729i$\\
			\hline\hline
		\end{tabular}
		\caption{Dominant quasinormal modes for the scalar perturbation with initial $\ell=1$ multipole. "-" means the duration of the ringdown phase is too short to give precise enough QNMs with Prony's method. }
		\label{QNMs}
	\end{table}	
	
	We notice that instability also occurs for other multipoles. In Fig. \ref{alphaPositiveL1m0Fig}, the time evolution of the multipole $\ell=1$  for various values of $\alpha>0$ are shown. We observe similar wave dynamics 
	behaviors due to the influence of $\alpha$ as the fundamental multipole mode. Besides, we learn that  $\alpha_c$ for $\ell=1$  is larger than its corresponding value for $\ell=0$.  
	
\begin{figure}[!htbp]
		\centering
		\subfigure[$~~a=0.3$ ]{\includegraphics[width=0.47\textwidth]{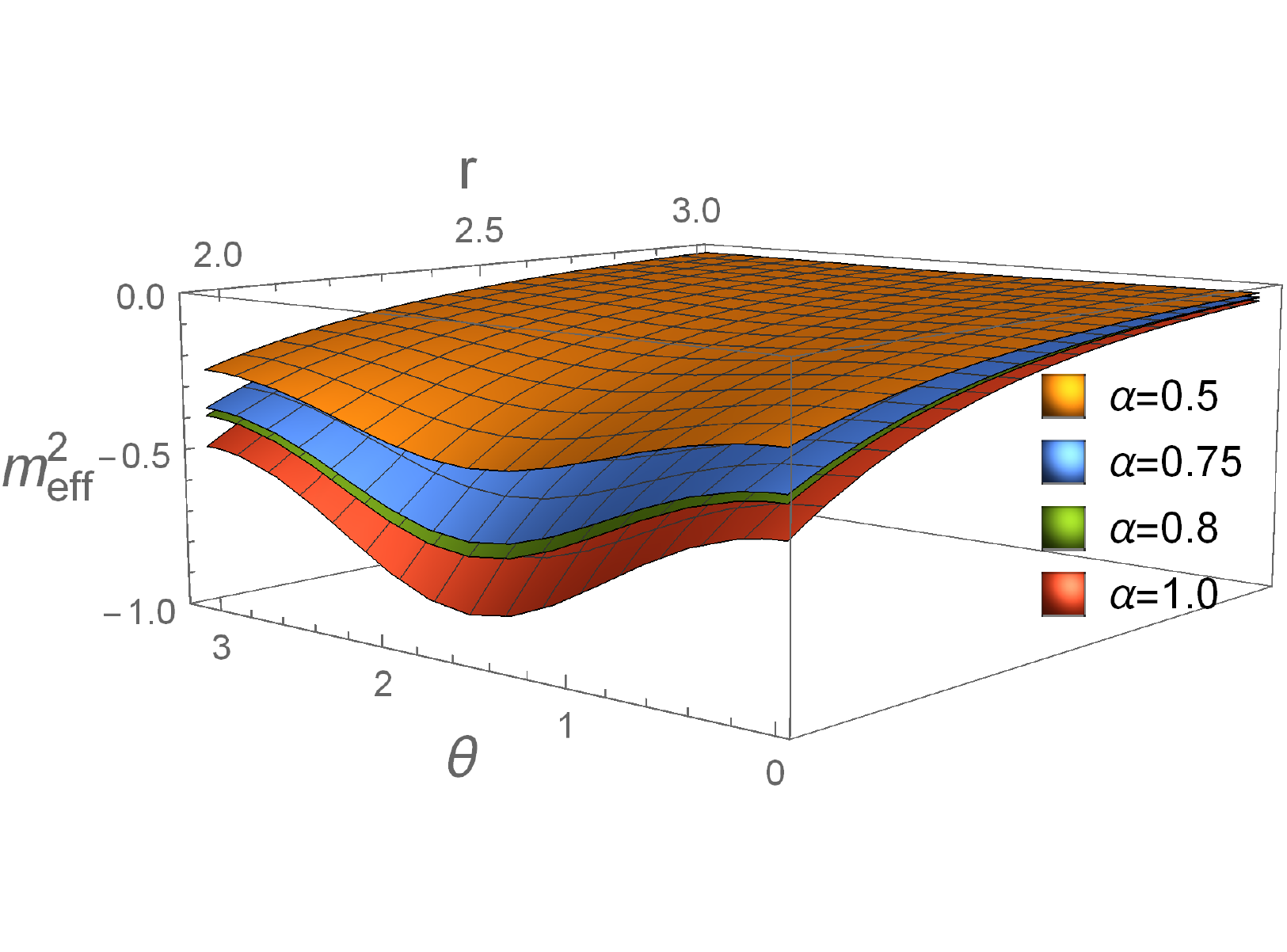}}\quad
		\subfigure[$~~a=0.9$ ]{\includegraphics[width=0.47\textwidth]{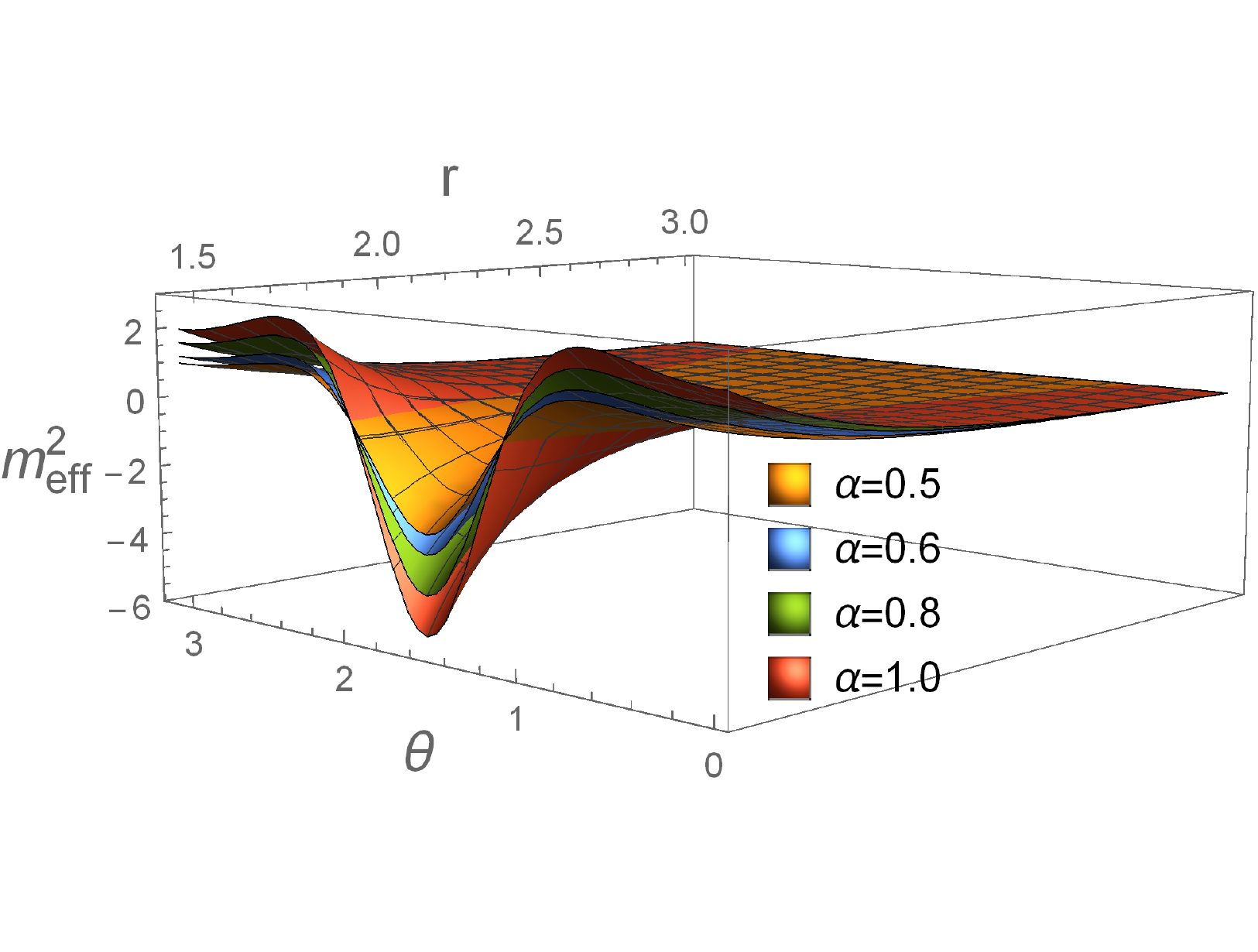}}
		\caption{(color online) The effective mass square $m^2_{\rm eff}$  is plotted for $a=0.3$ and $0.9$. $\alpha$ is chosen as in Fig. \ref{alphaPositiveL0m0Fig}.}
		\label{mSquarePositiveAlpha}
	\end{figure}

	To gain a better understanding of the physics behind the above phenomena, we examine the influence of the coupling constant and the spin on the effective mass square $m^2_{\rm eff}= -\alpha {\cal G}$. When $a=0$, the background reduces to the Schwarzschild spacetime in which ${\cal G} = 48 M^2/r^6$, so that $m^2_{\rm eff}$ is always negative everywhere for any $\alpha>0$ and   becomes more negative when  $\alpha$ is further increased. Small negative $m^2_{\rm eff}<0$ is not sufficient to trigger the tachyonic instability. Only when $m^2_{\rm eff}$ is sufficiently negative ($\alpha>\alpha_c$), can the instability be developed and becomes more violent with the further increase of $\alpha$, for which 
	  $m^2_{\rm eff}$ will become more negative. In Fig. \ref{mSquarePositiveAlpha}, $m^2_{\rm eff}$   is plotted for various $a \neq 0$ and $\alpha$. From the figure, one can clearly see that $m^2_{\rm eff}$ becomes negative somewhere even
	  when $\alpha$ is still smaller than the critical value. But this small negative effective mass square is not strong enough to change the bulk stability. As $\alpha$ continuously increases, at some fixed position close to the horizon, we see that $m^2_{\rm eff}$ becomes more and more negative.  The  instability for big enough $\alpha$ is triggered  by the negative enough $m^2_{\rm eff}$. Moreover, from the figure, one can see that with the increase of the spin, $m^2_{\rm eff}$ will be lifted up near the horizon in some $\theta$ direction. This cannot save the stability, since at other $\theta$ directions the effective mass square can become more negative, so that even smaller $\alpha_c$ can trigger the instability for fast rotating holes. 
	
	Now we examine the quasinormal ringing behavior. The influence of $\alpha$ on the ringdown phase is also explored by using the Prony's method, as shown in Table. \ref{QNMs}. Since the duration of the ringdown phase for the multipole  $\ell=0$ is too short to give precise enough QNMs by using Prony's method, we only present QNMs for  the $\ell=1$ multipole. From the table and the figure, one can see that when $a=0$, as $\alpha$ achieves  bigger positive values, the real part of QNMs $\Re \omega$ decreases,  while the imaginary part $|\Im \omega|$ increases. However, when $a \neq 0$, the situation becomes more complicated: as $\alpha$ increases from $0$,  $\Re \omega$ keeps decreasing for low spins but changes to increase for high spins, while  $|\Im \omega|$ first decreases and then increases. The non-monotonic change in $|\Im \omega|$ with the increase of the coupling constant becomes more obvious for fast rotation cases in QNMs. Since the QNMs  can be used  uniquely   to  identify  black
	holes (in GR) and shall be detected through GW observations (specially for the third generation detectors), it is clear that the fine structure in the QNMs due to the influence of the coupling between the scalar field and GB term  can be used to explore the signature of the sEGB theory.  Due to the  ``mode-mixing mechanism" \cite{Zenginoglu:2012us,Burko:2013bra,Thuestad:2017ngu}, one can expect that similar behaviors described above will also occur for higher multipoles.

	\subsection{$\alpha<0$}
	
	\begin{figure}[!htbp]
		\centering
		\subfigure[$~~a=0$ ]{\includegraphics[width=0.47\textwidth]{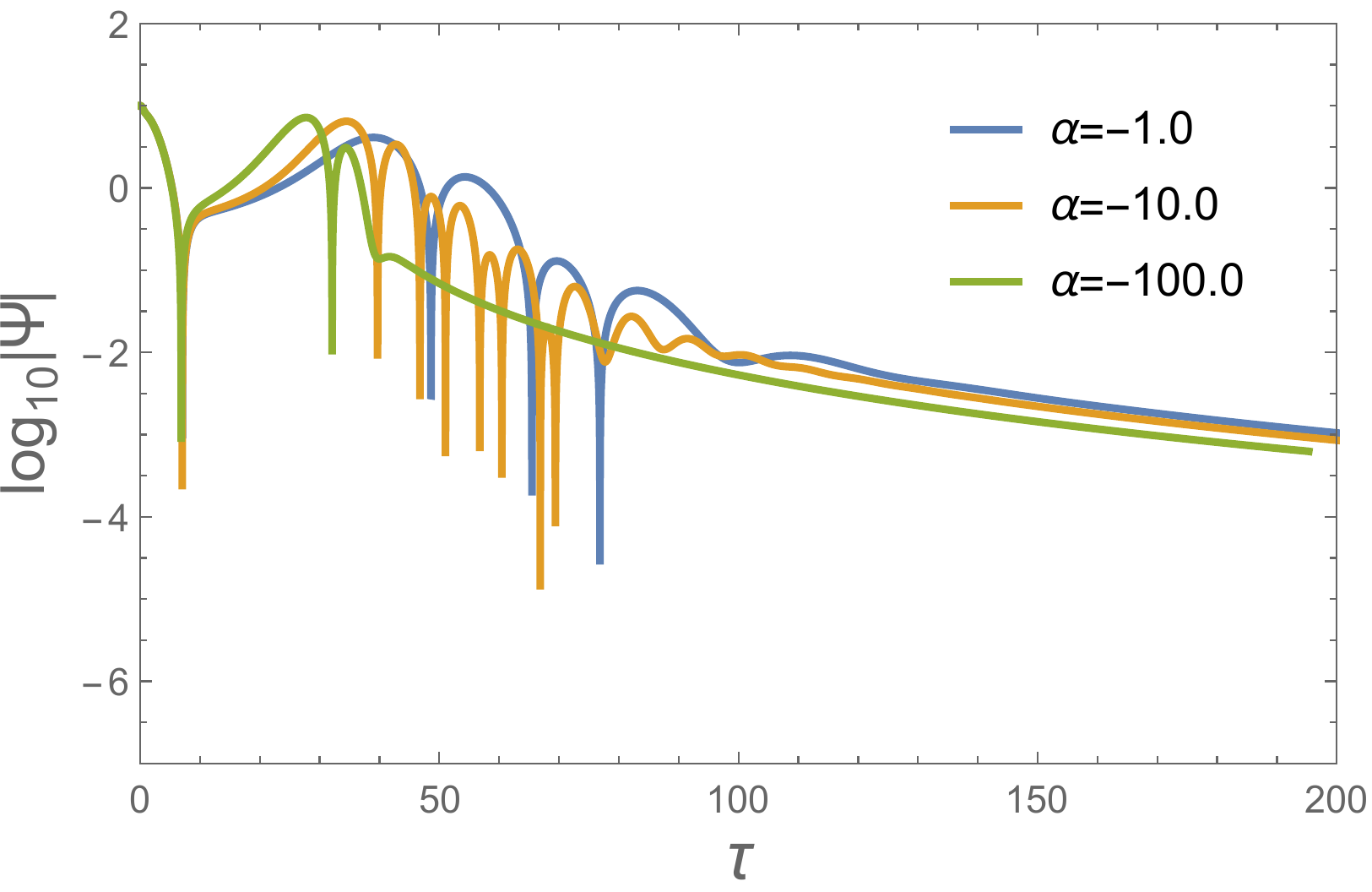}}\quad
		\subfigure[$~~a=0.5$ ]{\includegraphics[width=0.47\textwidth]{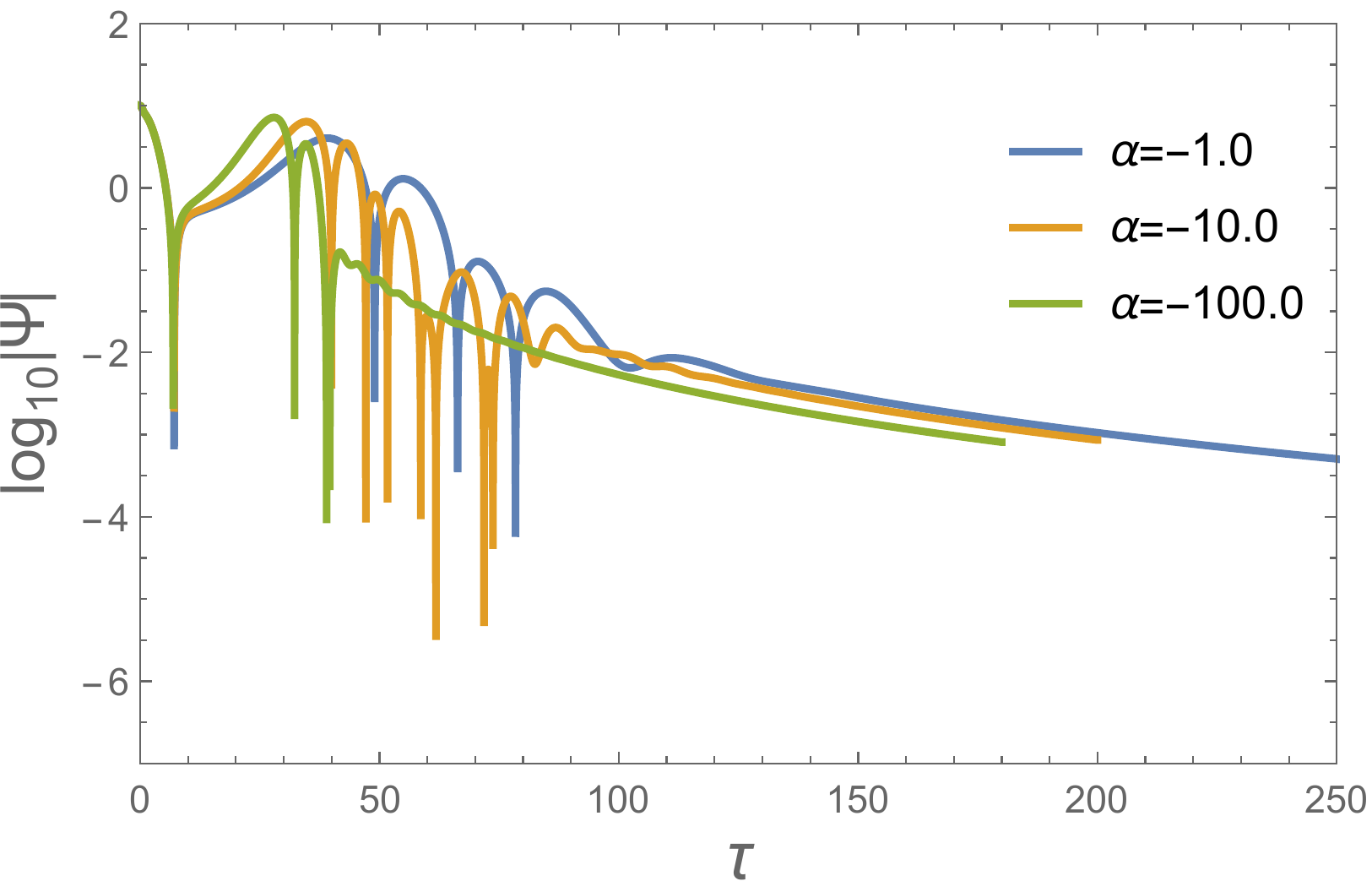}}
		\caption{(color online) Time evolution of the scalar perturbation for $a=0$ and $0.5$ with $\alpha<0$. Initial multipole is fixed as $\ell=0$. Time is in units of $M$.}
		\label{alphaNegativeL0m0Fig}
	\end{figure}
	
\begin{figure}[!htbp]
		\centering
		\subfigure[$~~a=0.7$ ]{\includegraphics[width=0.47\textwidth]{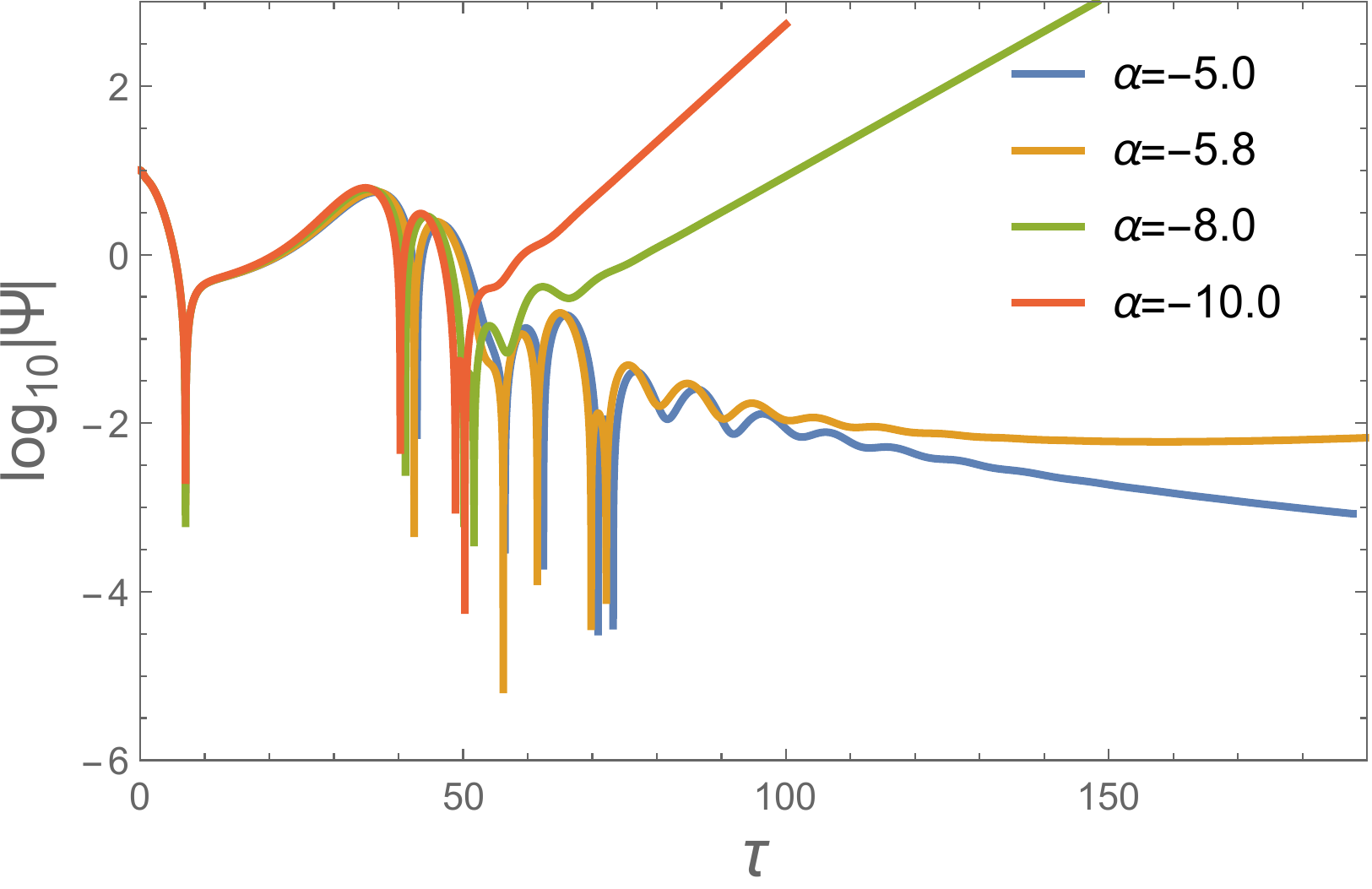}}\quad
		\subfigure[$~~a=0.8$ ]{\includegraphics[width=0.47\textwidth]{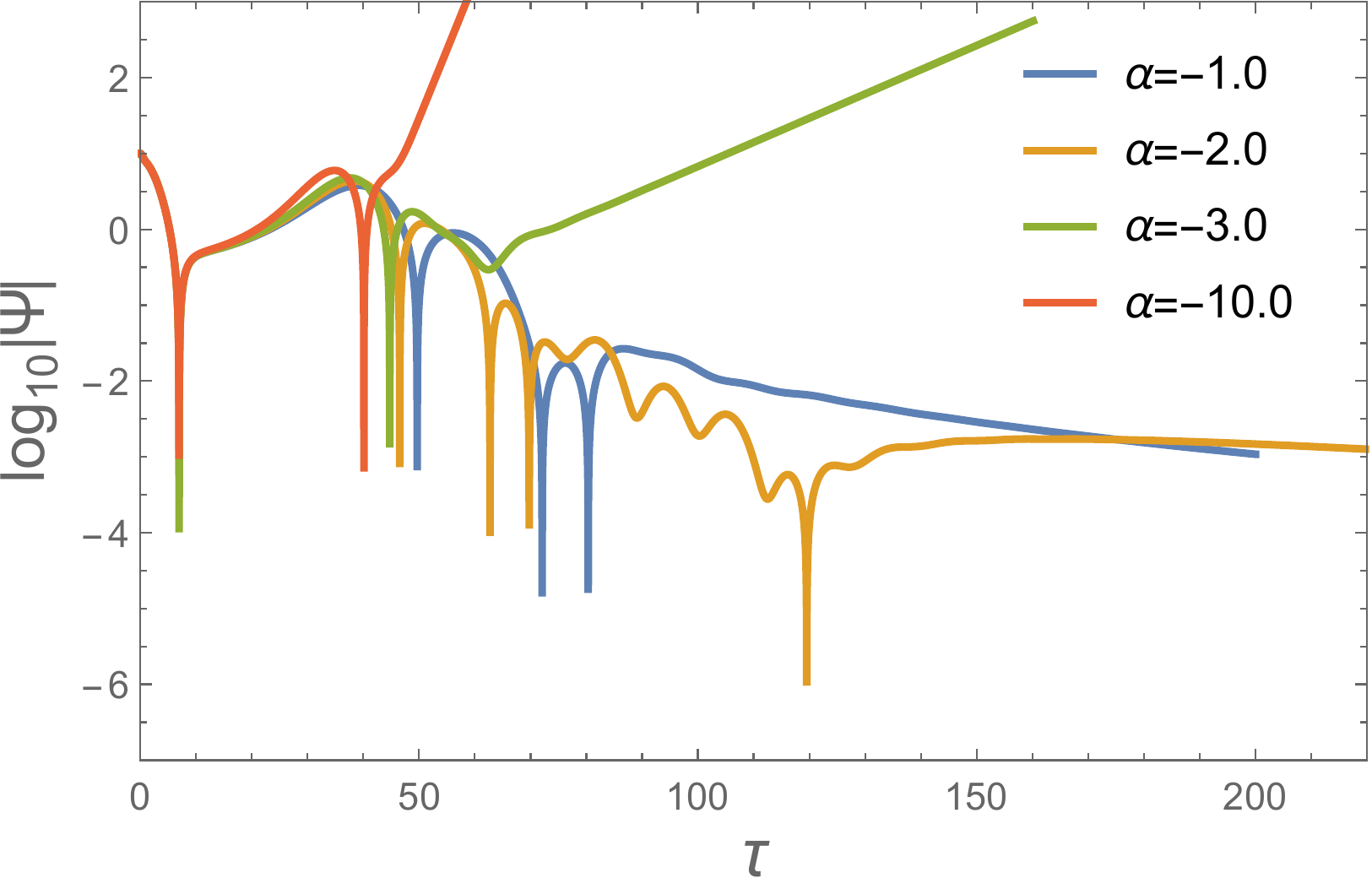}}\quad
		\subfigure[$~~a=0.9$ ]{\includegraphics[width=0.47\textwidth]{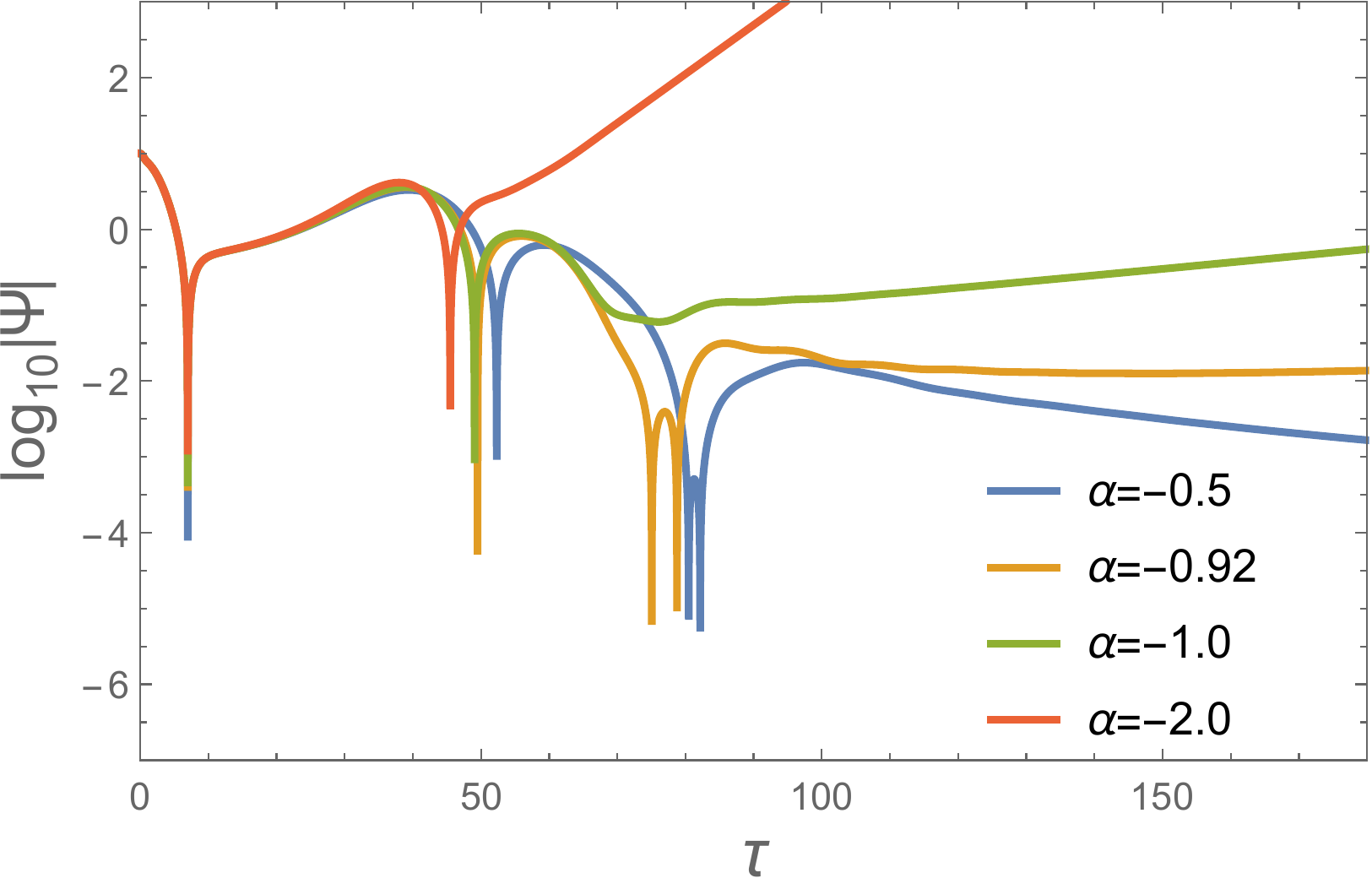}}
		\caption{(color online) Time evolution of the scalar perturbation for various spins. $a$ is fixed as $0.7, 0.8$ and $0.9$ respectively. Initial multipole is fixed as $\ell=0$. Time is in units of $M$.}
		\label{alphaNegativeL0m0Fig1}
	\end{figure}

	We now  consider the case with $\alpha<0$. In Fig. \ref{alphaNegativeL0m0Fig}, by fixing $a=0$ and $0.5$, the time evolution of the scalar perturbation for  multipole $\ell=0$ is plotted, where no instability is observed. This objective picture actually confirms  the argument that  there exists a minimum spin $a_{\rm min}=\frac{1}{2}$ and below which no instability can be triggered, which is irrespective of the value of $\alpha$  \cite{Dima:2020yac,Hod:2020jjy}.  To see more clearly  the influences of the spin and coupling constant on the onset of the instability, in Fig. \ref{alphaNegativeL0m0Fig1}, we fix the spin to some representative values above $a_{\rm min}$ and show the time evolution of the scalar perturbation for various $\alpha$.  From the figure, similar to the case $\alpha>0$, one can see that the instability only occurs when $\alpha$ exceeds a critical value $|\alpha_c|$ for a fixed spin. For  $a=0.7, 0,8$ and $0.9$, $\alpha_c$ is $-5.8,-2.0$ and $-0.92$,  respectively. This tells us that with the increase of the spin, $|\alpha_c|$ decreases quickly. For the extreme hole with the highest spin,  it is very simple to trigger instability because we only require the smallest $|\alpha_c|$. Moreover, we observe that above $|\alpha_c|$ for each chosen $a$, larger $|\alpha|$ makes the instability to appear earlier and behave more violently. Similar properties are also found for the multipole $\ell=1$.  These results agree well even quantitatively  with those found in \cite{Dima:2020yac,Hod:2020jjy,Doneva:2020nbb}.

	\begin{figure}[!htbp]
		\centering
		\subfigure[$~~a=0.7$ ]{\includegraphics[width=0.47\textwidth]{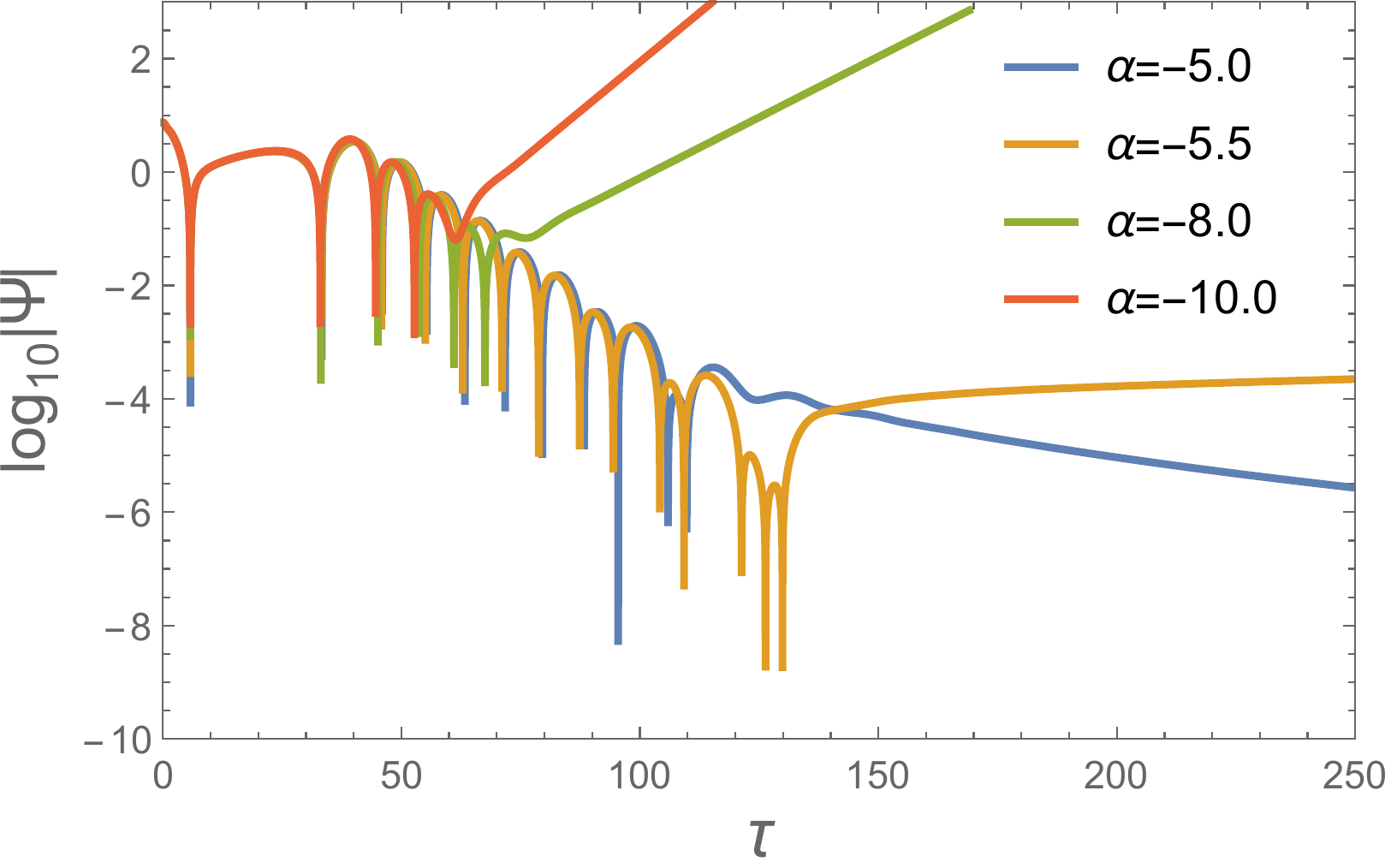}}\quad
		\subfigure[$~~a=0.8$ ]{\includegraphics[width=0.47\textwidth]{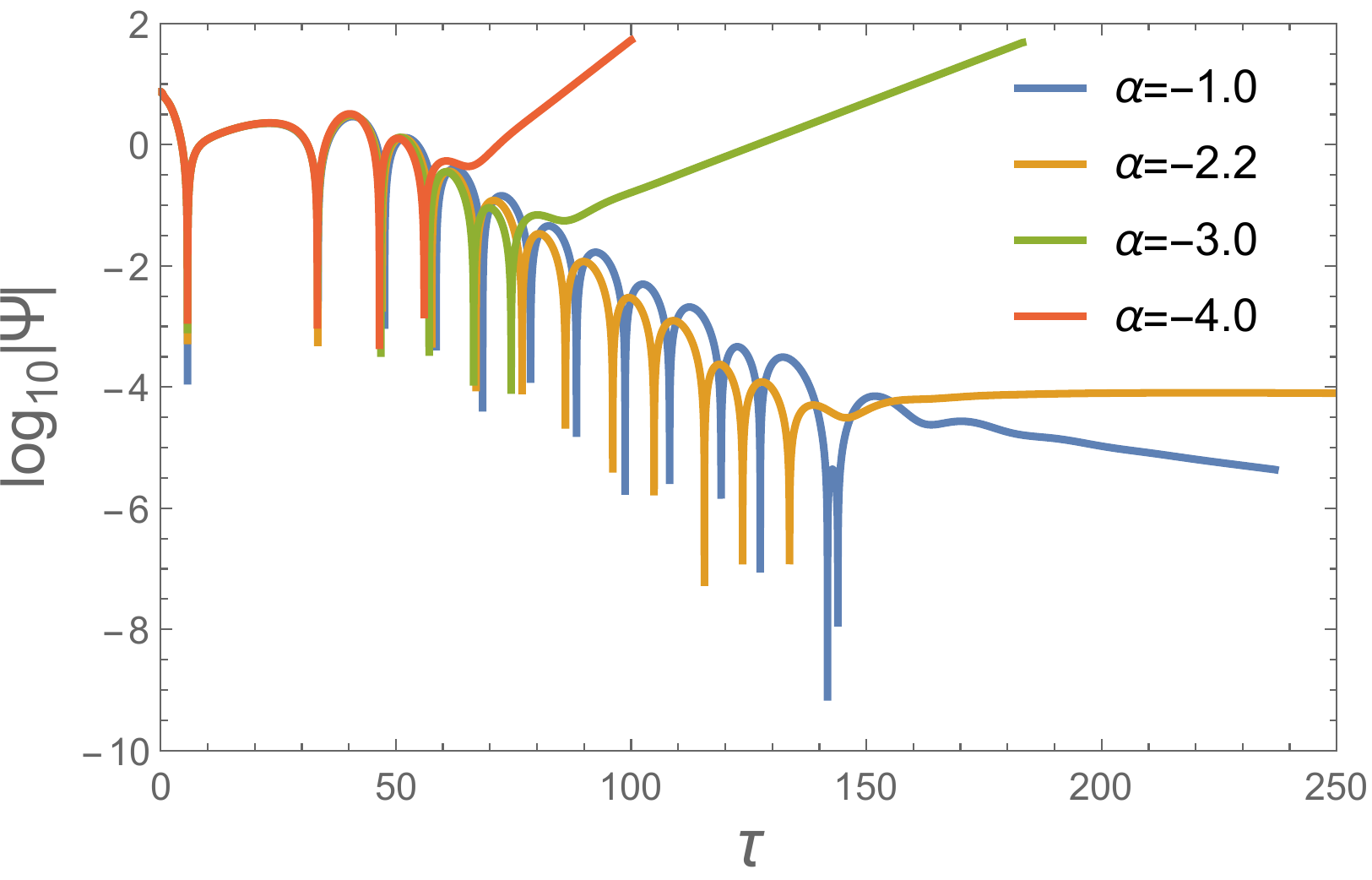}}\quad
		\subfigure[$~~a=0.9$ ]{\includegraphics[width=0.47\textwidth]{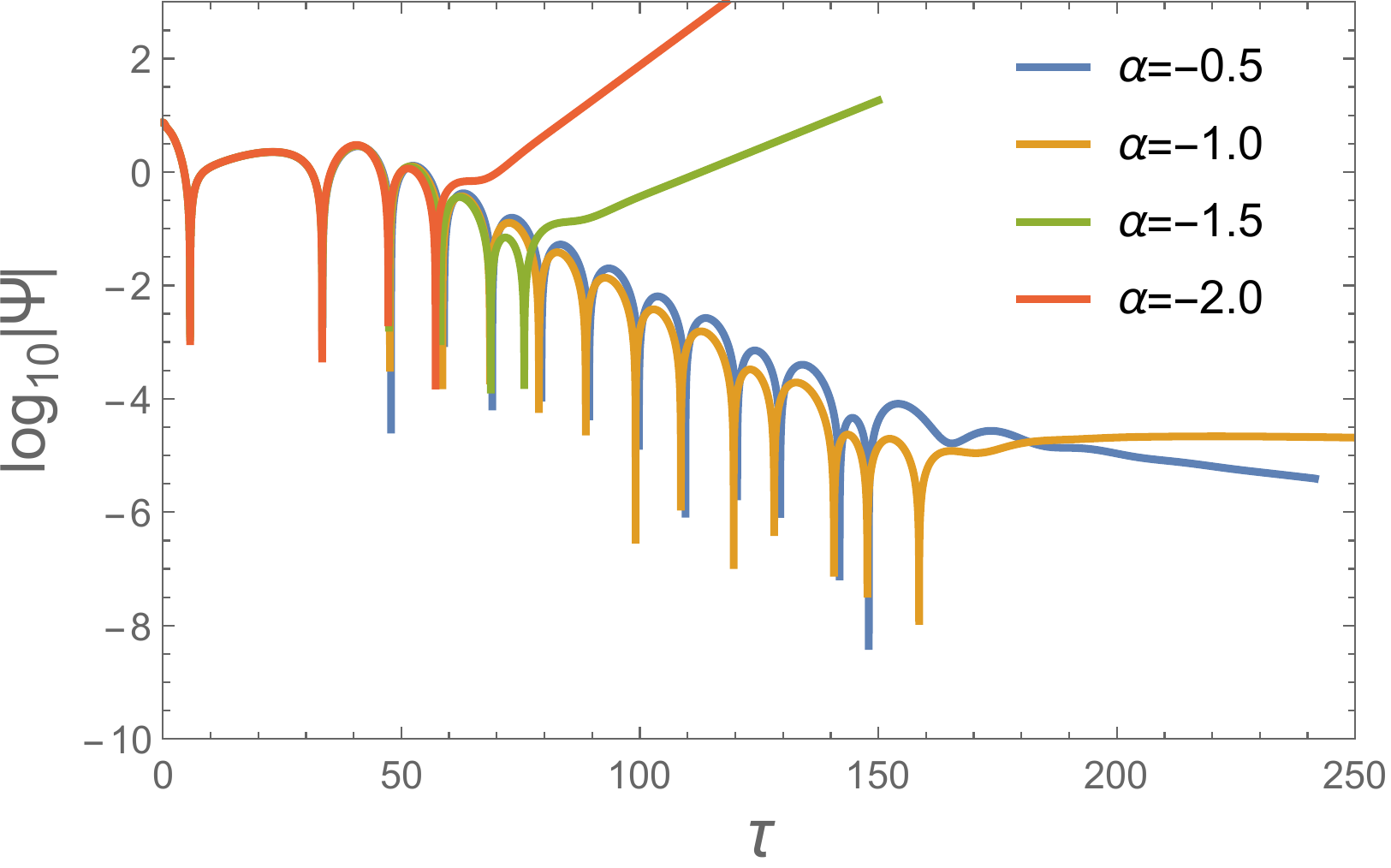}}
		\caption{(color online)  Time evolution of the scalar perturbation for various spins. $a$ is fixed as $0.7, 0.8$ and $0.9$ respectively. Initial multipole is fixed as $\ell=1$. Time is in units of $M$.}
		\label{alphaNegativeL1m0Fig}
	\end{figure}
	
	\begin{figure}[!htbp]
		\centering
		\subfigure[$~~a=0.5$ ]{\includegraphics[width=0.47\textwidth]{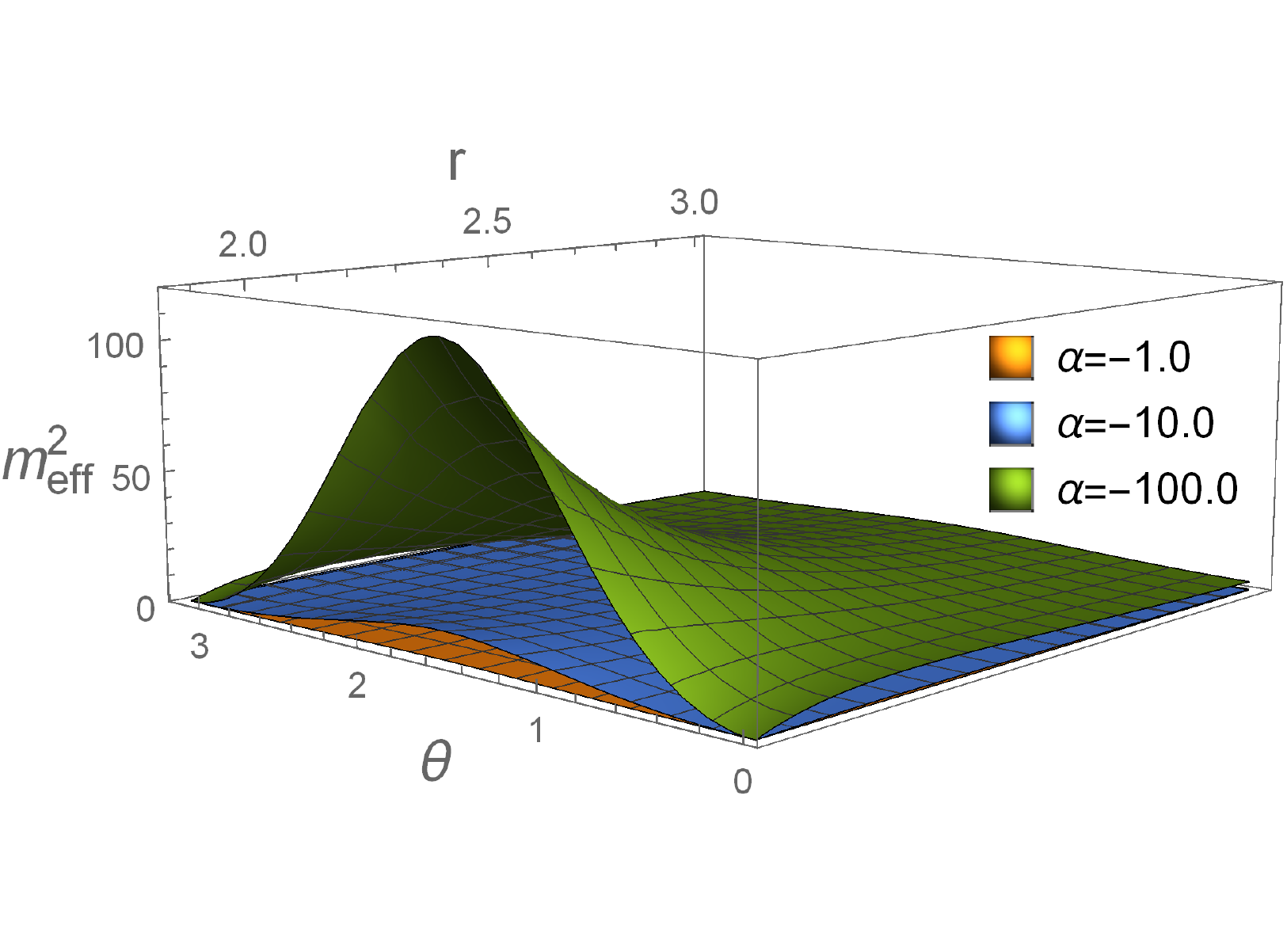}}\quad
		\subfigure[$~~a=0.9$ ]{\includegraphics[width=0.47\textwidth]{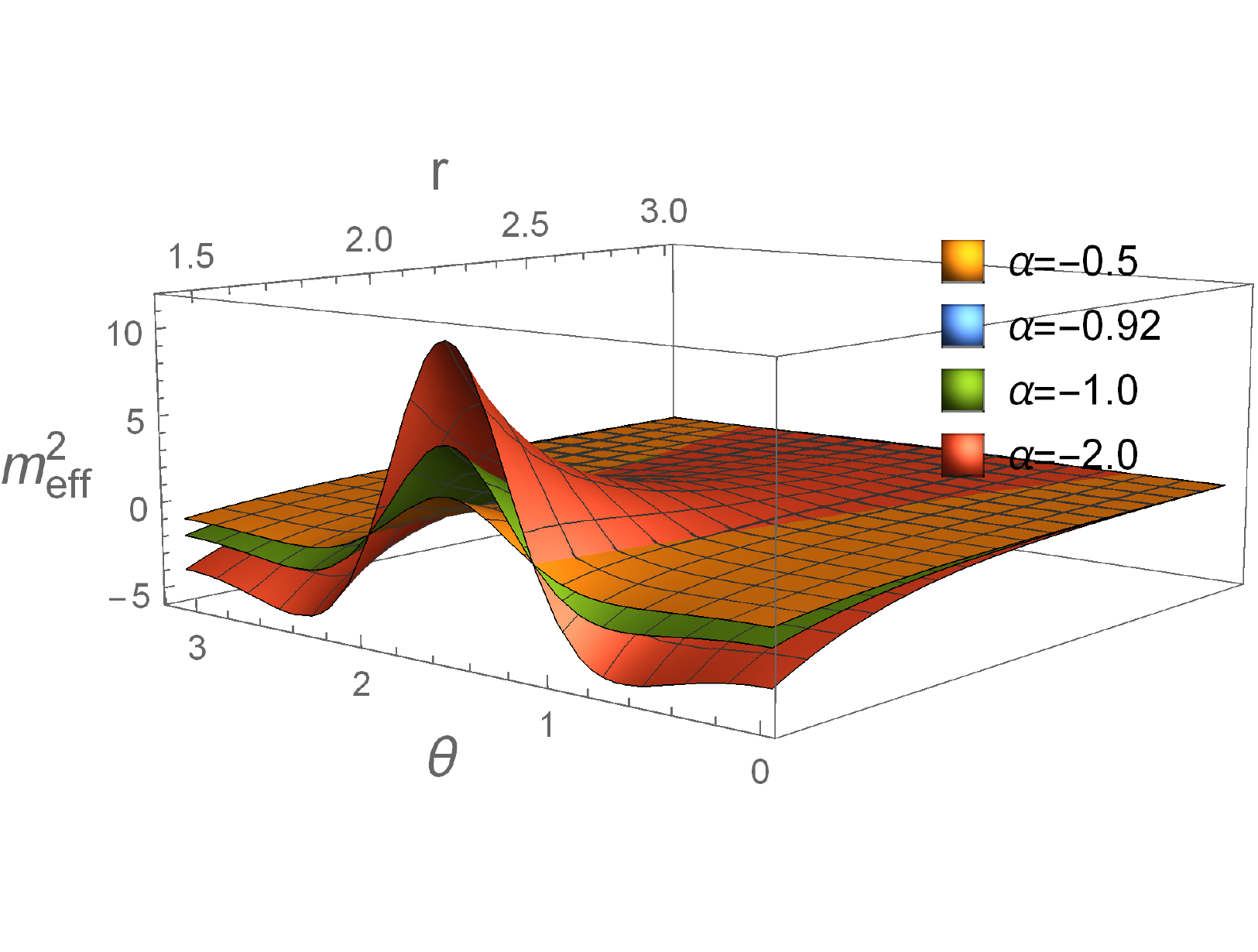}}
		\caption{(color online) The effective mass square $m^2_{\rm eff}$   is plotted for $a=0.5$ and $0.9$. $\alpha$ is chosen as in Figs. \ref{alphaNegativeL0m0Fig} and \ref{alphaNegativeL0m0Fig1}. In the right panel, the surface with $\alpha=-0.92$ overlaps largely with that with $\alpha=-1.0$.}
		\label{mSquareNegativeAlpha}
	\end{figure}
	
	Similar to the discussion for $\alpha>0$, we can also understand the instability from the behavior of the effective mass square $m^2_{\rm eff}$ and have a qualitative explanation. In Fig. \ref{mSquareNegativeAlpha}, $m^2_{\rm eff}$ as a function of the coordinates is plotted for various $\alpha<0$. From the left panel of the figure, one can see that when $a \leq a_{\rm min}=1/2$, $m^2_{\rm eff}$ is always positive everywhere for  any value of $\alpha<0$. This presents physically reason why there is no tachyonic instability for low spin $a \leq a_{\rm min}$. However, when $a>a_{\rm min}$, $a=0.9$ for example as shown in the right panel of the figure, we find that $m^2_{\rm eff}$ will become negative near the horizon, which explains the emergence of the tachyonic instability for sufficiently negative $\alpha$. Moreover, similar to the case with $\alpha>0$, as $|\alpha|$ is increased for a fixed spin,  $m^2_{\rm eff}$ becomes more and more negative at some fixed position, which explains the appearance of more and more violent instabilities.
	
	Besides the stability investigation in the final wave dynamics, we have used Prony's method and read off the QNMs in the time evolution of the perturbation in the ringdown phase for  $\ell=1$, see the results in Table \ref{QNMs}. We find that the result is simpler than the case with $\alpha>0$ mentioned above.  For a chosen $a$, with the decrease of $|\alpha|$, both $\Re \omega$ and $|\Im \omega|$ decrease. Differences in QNM frequencies from the GR situation show that in sEGB theory the characteristic sound is special. We expect that future precise detections of  GWs can observe the signature of the sEGB theory from its special sound frequency. Once again, due to the ``mode-mixing mechanism" \cite{Zenginoglu:2012us,Burko:2013bra,Thuestad:2017ngu}, one can expect that instability will also occur for higher  multipoles, and moreover even and odd modes will give roughly comparable contributions to the instability \cite{Dima:2020yac}.

	\section{Summary and Discussions}
	
	In this work, we have studied the massless scalar perturbation on the background of the Kerr black hole in sEGB theory. We have overcome the nonseparability problem and the ``outer boundary problem" by employing the hyperboloidal compactification technique to perform carefully the numerical computations and have got object pictures on the time evolution of the scalar field  perturbations. In addition to the information about the stability, we have also obtained the sound characteristics of the quasinormal ringing in the sEGB theory. For negative couplings between the scalar field and the GB term, we have presented objective support on the existence of the minimum rotation to break the Kerr stability, no matter how strong the coupling $\alpha$ is. This minimum $a$ requirement does not exist for positive $\alpha$s. For  fast rotating black holes, it requires smaller coupling $|\alpha|$ to trigger 
	the instability. This result holds for negative couplings when the rotation is above a minimum value.  The stronger coupling $|\alpha|$ will bring the break down of the original Kerr configuration to happen earlier and the resulting instability becomes more violent.  We have also examined carefully the characteristic sound of the quasinormal ringing of the scalar perturbation outside the original Kerr black hole in the sEGB theory. We have unveiled the special sound once there is a coupling between the scalar field and the GB curvature and when it becomes stronger. The fine frequency difference uncovered can be used  as imprints to confirm the sEGB theory in the future precise GW detections.
	
	In addition to presenting objective pictures of the wave propagation in the sEGB theory, we have also examined the physical reason about the occurrence of the instability of the Kerr black hole once the coupling between the scalar field and GB term is strong enough. We have found that a big enough $|\alpha|$ always leads to a negative  effective   mass square, which triggers the  tachyonic instability and destroys the original Kerr background configuration in the sEGB theory.  With the increase of $|\alpha|$, the effective mass square  becomes more negative, which results in a more violent instability. For negative couplings, we have found the reason behind the minimum rotation in order to accommodate the instability, and the effective mass square is always positive no matter how strong  the coupling strength is.  
	
    In this paper, we have concentrated ourselves on the massless scalar field perturbation. It is of great interest to generalize the discussion to the massive scalar field. For the massive scalar field, in addition to the tachyonic instability, there is the possibility for the superradiant instability to happen. There will be richer physics involved in studying the dynamics and the fate of the Kerr black hole. Furthermore, in this study we have only paid attention to the probe limit, the linear perturbation in the  Kerr black hole background. It is intriguing to extend our study to include the backreaction of the scalar field, which can present us the process of the formation of the scalar hair and the final spacetime of the hairy black hole.  Another interesting direction is to extend our study to consider the coupling of the scalar field to other topological curvature invariants beyond the Gauss-Bonnet term we considered here, such as Chern-Simons term \cite{Gao:2018acg}, Lovelock term, etc.

	\begin{acknowledgments}
		
		This work is partially supported by the National Natural Science Foundation of China (NNSFC) under Grant Nos. 12075202, 11675145,  11975203, and 12075207. J.S. is supported by ANID Chile through FONDECYT Grant  No.1170279.
	\end{acknowledgments}


\begin{thebibliography}{99}
		
		\bibitem{Abbott:2016blz}
		B.~Abbott \textit{et al.} [LIGO Scientific and Virgo],
		{\em Observation of Gravitational Waves from a Binary Black Hole Merger},
		Phys. Rev. Lett. \textbf{116}, no.6, 061102 (2016)
		[arXiv:1602.03837 [gr-qc]].
		
		\bibitem{Abbott:2016nmj}
		B.~P.~Abbott \textit{et al.} [LIGO Scientific and Virgo],
		{\em GW151226: Observation of Gravitational Waves from a 22-Solar-Mass Binary Black Hole Coalescence},
		Phys. Rev. Lett. \textbf{116}, no.24, 241103 (2016)
		[arXiv:1606.04855 [gr-qc]].
		
		\bibitem{Abbott:2017gyy}
		B.~P.~Abbott \textit{et al.} [LIGO Scientific and Virgo],
		{\em GW170608: Observation of a 19-solar-mass Binary Black Hole Coalescence},
		Astrophys. J. \textbf{851}, no.2, L35 (2017)
		doi:10.3847/2041-8213/aa9f0c
		[arXiv:1711.05578 [astro-ph.HE]].
		
		\bibitem{Akiyama:2019cqa}
		K.~Akiyama \textit{et al.} [Event Horizon Telescope],
		{\em First M87 Event Horizon Telescope Results. I. The Shadow of the Supermassive Black Hole},
		Astrophys. J. \textbf{875}, no.1, L1 (2019)
		[arXiv:1906.11238 [astro-ph.GA]].
		
		\bibitem{Akiyama:2019eap}
		K.~Akiyama \textit{et al.} [Event Horizon Telescope],
		{\em First M87 Event Horizon Telescope Results. VI. The Shadow and Mass of the Central Black Hole},
		Astrophys. J. Lett. \textbf{875}, no.1, L6 (2019)
		[arXiv:1906.11243 [astro-ph.GA]].
		
		\bibitem{Carter:1971zc}
		B.~Carter,
		{\em Axisymmetric Black Hole Has Only Two Degrees of Freedom},
		Phys. Rev. Lett. \textbf{26}, 331-333 (1971)
		
		\bibitem{Robinson:1975bv}
		D.~Robinson,
		{\em Uniqueness of the Kerr black hole},
		Phys. Rev. Lett. \textbf{34}, 905-906 (1975)
		
		\bibitem{Chrusciel:2012jk}
		P.~T.~Chrusciel, J.~Lopes Costa and M.~Heusler,
		{\em Stationary Black Holes: Uniqueness and Beyond},
		Living Rev. Rel. \textbf{15}, 7 (2012)
		[arXiv:1205.6112 [gr-qc]].
		
		\bibitem{Sotiriou:2015pka}
		T.~P.~Sotiriou,
		{\em Black Holes and Scalar Fields},
		Class. Quant. Grav. \textbf{32}, no.21, 214002 (2015)
		[arXiv:1505.00248 [gr-qc]].
		
		\bibitem{Herdeiro:2015waa}
		C.~A.~Herdeiro and E.~Radu,
		{\em Asymptotically flat black holes with scalar hair: a review},
		Int. J. Mod. Phys. D \textbf{24}, no.09, 1542014 (2015)
		[arXiv:1504.08209 [gr-qc]].
		
		\bibitem{Berti:2018cxi}
		E.~Berti, K.~Yagi and N.~Yunes,
		{\em Extreme Gravity Tests with Gravitational Waves from Compact Binary Coalescences: (I) Inspiral-Merger},
		Gen. Rel. Grav. \textbf{50}, no.4, 46 (2018)
		[arXiv:1801.03208 [gr-qc]].
		
		
		\bibitem{Doneva:2017bvd}
		D.~D.~Doneva and S.~S.~Yazadjiev,
		{\em New Gauss-Bonnet Black Holes with Curvature-Induced Scalarization in Extended Scalar-Tensor Theories},
		Phys. Rev. Lett. \textbf{120}, no.13, 131103 (2018)
		[arXiv:1711.01187 [gr-qc]].
		
		\bibitem{Silva:2017uqg}
		H.~O.~Silva, J.~Sakstein, L.~Gualtieri, T.~P.~Sotiriou and E.~Berti,
		{\em Spontaneous scalarization of black holes and compact stars from a Gauss-Bonnet coupling},
		Phys. Rev. Lett. \textbf{120}, no.13, 131104 (2018)
		[arXiv:1711.02080 [gr-qc]].
		
		\bibitem{Antoniou:2017acq}
		G.~Antoniou, A.~Bakopoulos and P.~Kanti,
		{\em Evasion of No-Hair Theorems and Novel Black-Hole Solutions in Gauss-Bonnet Theories},
		Phys. Rev. Lett. \textbf{120}, no.13, 131102 (2018)
		[arXiv:1711.03390 [hep-th]].
		
		\bibitem{Cunha:2019dwb}
		P.~V.~Cunha, C.~A.~Herdeiro and E.~Radu,
		{\em Spontaneously Scalarized Kerr Black Holes in Extended Scalar-Tensor–Gauss-Bonnet Gravity},
		Phys. Rev. Lett. \textbf{123}, no.1, 011101 (2019)
		[arXiv:1904.09997 [gr-qc]].
		
		\bibitem{Herdeiro:2020wei}
		C.~A.~R.~Herdeiro, E.~Radu, H.~O.~Silva, T.~P.~Sotiriou and N.~Yunes,
		{\em Spin-induced scalarized black holes},
		[arXiv:2009.03904 [gr-qc]].
		
		\bibitem{Berti:2020kgk}
		E.~Berti, L.~G.~Collodel, B.~Kleihaus and J.~Kunz,
		{\em Spin-induced black-hole scalarization in Einstein-scalar-Gauss-Bonnet theory},
		[arXiv:2009.03905 [gr-qc]].
		
		\bibitem{Damour:1993hw}
		T.~Damour and G.~Esposito-Farese,
		{\em Nonperturbative strong field effects in tensor - scalar theories of gravitation},
		Phys. Rev. Lett. \textbf{70}, 2220-2223 (1993)
		
		
		
		\bibitem{Blazquez-Salcedo:2018jnn}
		J.~L.~Blázquez-Salcedo, D.~D.~Doneva, J.~Kunz and S.~S.~Yazadjiev,
		{\em Radial perturbations of the scalarized Einstein-Gauss-Bonnet black holes},
		Phys. Rev. D \textbf{98}, no.8, 084011 (2018)
		[arXiv:1805.05755 [gr-qc]].
		
		\bibitem{Silva:2018qhn}
		H.~O.~Silva, C.~F.~B.~Macedo, T.~P.~Sotiriou, L.~Gualtieri, J.~Sakstein and E.~Berti,
		{\em Stability of scalarized black hole solutions in scalar-Gauss-Bonnet gravity},
		Phys. Rev. D \textbf{99}, no.6, 064011 (2019)
		[arXiv:1812.05590 [gr-qc]].
		
		\bibitem{Vishveshwara:1970zz}
		C.~V.~Vishveshwara,
	    {\em Scattering of Gravitational Radiation by a Schwarzschild Black-hole},
		Nature \textbf{227}, 936-938 (1970)
		
		
		\bibitem{Kokkotas:1999bd}
		K.~D.~Kokkotas and B.~G.~Schmidt,
		{\em Quasinormal modes of stars and black holes},
		Living Rev. Rel. \textbf{2}, 2 (1999)
		[arXiv:gr-qc/9909058 [gr-qc]].
		
		\bibitem{Nollert:1999ji}
		H.~P.~Nollert,
		{\em TOPICAL REVIEW: Quasinormal modes: the characteristic `sound' of black holes and neutron stars},
		Class. Quant. Grav. \textbf{16}, R159-R216 (1999)
		
		\bibitem{Berti:2009kk}
		E.~Berti, V.~Cardoso and A.~O.~Starinets,
		{\em Quasinormal modes of black holes and black branes},
		Class. Quant. Grav. \textbf{26}, 163001 (2009)
		[arXiv:0905.2975 [gr-qc]].
		
		\bibitem{Konoplya:2011qq}
		R.~Konoplya and A.~Zhidenko,
		{\em Quasinormal modes of black holes: From astrophysics to string theory},
		Rev. Mod. Phys. \textbf{83}, 793-836 (2011)
		[arXiv:1102.4014 [gr-qc]].
		
		\bibitem{Berti:2018vdi}
		E.~Berti, K.~Yagi, H.~Yang and N.~Yunes,
		{\em Extreme Gravity Tests with Gravitational Waves from Compact Binary Coalescences: (II) Ringdown},
		Gen. Rel. Grav. \textbf{50}, no.5, 49 (2018)
		[arXiv:1801.03587 [gr-qc]].
		
		
		\bibitem{Krivan:1996da}
		W.~Krivan, P.~Laguna and P.~Papadopoulos,
		{\em Dynamics of scalar fields in the background of rotating black holes},
		Phys. Rev. D \textbf{54}, 4728-4734 (1996)
		[arXiv:gr-qc/9606003 [gr-qc]].
		
		\bibitem{PazosAvalos:2004rp}
		E.~Pazos-Avalos and C.~O.~Lousto,
		{\em Numerical integration of the Teukolsky equation in the time domain},
		Phys. Rev. D \textbf{72}, 084022 (2005)
		[arXiv:gr-qc/0409065 [gr-qc]].
		
		\bibitem{Dolan:2011dx}
		S.~R.~Dolan, L.~Barack and B.~Wardell,
		{\em Self force via $m$-mode regularization and 2+1D evolution: II. Scalar-field implementation on Kerr spacetime},
		Phys. Rev. D \textbf{84}, 084001 (2011)
		[arXiv:1107.0012 [gr-qc]].
		
		\bibitem{Thuestad:2017ngu}
		I.~Thuestad, G.~Khanna and R.~H.~Price,
		{\em Scalar Fields in Black Hole Spacetimes},
		Phys. Rev. D \textbf{96}, no.2, 024020 (2017)
		[arXiv:1705.04949 [gr-qc]].
		
		
		
		
		\bibitem{Dolan:2012yt}
		S.~R.~Dolan,
		{\em Superradiant instabilities of rotating black holes in the time domain},
		Phys. Rev. D \textbf{87}, no.12, 124026 (2013)
		[arXiv:1212.1477 [gr-qc]].
		
		\bibitem{Brito:2014nja}
		R.~Brito, V.~Cardoso and P.~Pani,
		{\em Superradiant instability of black holes immersed in a magnetic field},
		Phys. Rev. D \textbf{89}, no.10, 104045 (2014)
		[arXiv:1405.2098 [gr-qc]].
		
		\bibitem{Zenginoglu:2007jw}
		A.~Zenginoglu,
		{\em Hyperboloidal foliations and scri-fixing},
		Class. Quant. Grav. \textbf{25}, 145002 (2008)
		[arXiv:0712.4333 [gr-qc]].
		
		\bibitem{Zenginoglu:2008wc}
		A.~Zenginoglu,
		{\em A Hyperboloidal study of tail decay rates for scalar and Yang-Mills fields},
		Class. Quant. Grav. \textbf{25}, 175013 (2008)
		[arXiv:0803.2018 [gr-qc]].
		
		\bibitem{Zenginoglu:2008pw}
		A.~Zenginoglu,
		{\em Hyperboloidal evolution with the Einstein equations},
		Class. Quant. Grav. \textbf{25}, 195025 (2008)
		[arXiv:0808.0810 [gr-qc]].
		
		\bibitem{Zenginoglu:2008uc}
		A.~Zenginoglu, D.~Nunez and S.~Husa,
		{\em Gravitational perturbations of Schwarzschild spacetime at null infinity and the hyperboloidal initial value problem},
		Class. Quant. Grav. \textbf{26}, 035009 (2009)
		[arXiv:0810.1929 [gr-qc]].
		
		\bibitem{Zenginoglu:2009hd}
		A.~Zenginoglu and M.~Tiglio,
		{\em Spacelike matching to null infinity},
		Phys. Rev. D \textbf{80}, 024044 (2009)
		[arXiv:0906.3342 [gr-qc]].
		
		\bibitem{Zenginoglu:2009ey}
		A.~Zenginoglu,
		{\em Asymptotics of black hole perturbations},
		Class. Quant. Grav. \textbf{27}, 045015 (2010)
		[arXiv:0911.2450 [gr-qc]].
		
		\bibitem{Zenginoglu:2010cq}
		A.~Zenginoglu,
		{\em Hyperboloidal layers for hyperbolic equations on unbounded domains},
		J. Comput. Phys. \textbf{230}, 2286-2302 (2011)
		[arXiv:1008.3809 [math.NA]].
		
		\bibitem{Zenginoglu:2011jz}
		A.~Zenginoglu,
		{\em A Geometric framework for black hole perturbations},
		Phys. Rev. D \textbf{83}, 127502 (2011)
		[arXiv:1102.2451 [gr-qc]].
		
		\bibitem{Racz:2011qu}
		I.~Racz and G.~Z.~Toth,
		{\em Numerical investigation of the late-time Kerr tails},
		Class. Quant. Grav. \textbf{28}, 195003 (2011)
		[arXiv:1104.4199 [gr-qc]].
		
		\bibitem{Zenginoglu:2011zz}
		A.~Zenginoglu and G.~Khanna,
		{\em Null infinity waveforms from extreme-mass-ratio inspirals in Kerr spacetime},
		Phys. Rev. X \textbf{1}, 021017 (2011)
		[arXiv:1108.1816 [gr-qc]].
		
		\bibitem{Cano:2020cao}
		P.~A.~Cano, K.~Fransen and T.~Hertog,
		{\em Ringing of rotating black holes in higher-derivative gravity},
		Phys. Rev. D \textbf{102}, no.4, 044047 (2020)
		[arXiv:2005.03671 [gr-qc]].
		

		
		\bibitem{Dima:2020yac}
		A.~Dima, E.~Barausse, N.~Franchini and T.~P.~Sotiriou,
		{\em Spin-induced black hole spontaneous scalarization},
		[arXiv:2006.03095 [gr-qc]].
		
		\bibitem{Hod:2020jjy}
		S.~Hod,
		{\em Onset of spontaneous scalarization in spinning Gauss-Bonnet black holes},
		[arXiv:2006.09399 [gr-qc]].
		
		\bibitem{Doneva:2020nbb}
		D.~D.~Doneva, L.~G.~Collodel, C.~J.~Krüger and S.~S.~Yazadjiev,
		{\em Black hole scalarization induced by the spin -- 2+1 time evolution},
		[arXiv:2008.07391 [gr-qc]].
		
		
		\bibitem{Konoplya:2019hml}
		R.~A.~Konoplya, A.~F.~Zinhailo and Z.~Stuchl\'\i{}k,
		{\em Quasinormal modes, scattering, and Hawking radiation in the vicinity of an Einstein-dilaton-Gauss-Bonnet black hole},
		Phys. Rev. D \textbf{99}, no.12, 124042 (2019)
		[arXiv:1903.03483 [gr-qc]].
		
		\bibitem{Zenginoglu:2012us}
		A.~Zenginoğlu, G.~Khanna and L.~M.~Burko,
		{\em Intermediate behavior of Kerr tails},
		Gen. Rel. Grav. \textbf{46}, 1672 (2014)
		[arXiv:1208.5839 [gr-qc]].
		
		\bibitem{Burko:2013bra}
		L.~M.~Burko and G.~Khanna,
		{\em Mode coupling mechanism for late-time Kerr tails},
		Phys. Rev. D \textbf{89}, no.4, 044037 (2014)
		[arXiv:1312.5247 [gr-qc]].
		
		
		
		\bibitem{Gao:2018acg}
		Y.~Gao, Y.~Huang and D.~Liu,
		{\em Scalar perturbations on the background of Kerr black holes in the quadratic dynamical Chern-Simons gravity},
		Phys.\ Rev.\ D \textbf{99}, no.4, 044020 (2019)
		[arXiv:1808.01433 [gr-qc]].
		
		\bibitem{Harms:2014dqa}
		E.~Harms, S.~Bernuzzi, A.~Nagar and A.~Zenginoglu,
		{\em A new gravitational wave generation algorithm for particle perturbations of the Kerr spacetime},
		Class. Quant. Grav. \textbf{31}, no.24, 245004 (2014)
		[arXiv:1406.5983 [gr-qc]].
		
		\bibitem{Schiesser}
		W.E. Schiesser, 
		{\em The Numerical Method of Lines: Integration of Partial Differential Equations}
		(Academic Press, New York, 1991).
		
		\bibitem{Berti:2007dg}
		E.~Berti, V.~Cardoso, J.~A.~Gonzalez and U.~Sperhake,
		{\em Mining information from binary black hole mergers: A Comparison of estimation methods for complex exponentials in noise},
		Phys. Rev. D \textbf{75}, 124017 (2007)
		[arXiv:gr-qc/0701086 [gr-qc]].
		
		\bibitem{Dolan:2007mj}
		S.~R.~Dolan,
		{\em Instability of the massive Klein-Gordon field on the Kerr spacetime},
		Phys. Rev. D \textbf{76}, 084001 (2007)
		[arXiv:0705.2880 [gr-qc]].
		
		
	\end{thebibliography}
\end{document}